\documentclass[%
 reprint,
 amsmath,amssymb,
 aps,
]{revtex4-1}
\usepackage{url}
\usepackage[colorlinks,linkcolor=blue]{hyperref}
\usepackage{graphicx}
\usepackage{dcolumn}
\usepackage{bm}

\begin{document}

\title{Two-path interference for enantiomer-selective state transfer of chiral molecules}
\author{Jin-Lei Wu$^{1}$}\author{Yan Wang$^{1}$}\author{Jin-Xuan Han$^{1}$}\author{Cong Wang$^{1}$}\author{Shi-Lei Su$^{2}$}\author{Yan Xia$^{3}$}\author{Yongyuan Jiang$^{1}$}\email[]{jiangyy@hit.edu.cn}\author{Jie Song$^{1,4,5}$}\email[]{jsong@hit.edu.cn}
\affiliation{$^{1}$School of Physics, Harbin Institute of Technology, Harbin 150001, China\\
$^{2}$School of Physics, Zhengzhou University, Zhengzhou 450001, China\\
$^{3}$Department of Physics, Fuzhou University, Fuzhou 350002, China\\
$^{4}$Key Laboratory of Micro-Nano Optoelectronic Information System, Ministry of Industry and Information Technology, Harbin 150001, China\\
$^{5}$Key Laboratory of Micro-Optics and Photonic Technology of Heilongjiang Province, Harbin Institute of Technology, Harbin 150001, China}

\begin{abstract}
With a microwave-regime cyclic three-state configuration, an enantiomer-selective state transfer~(ESST) is carried out through the two-path interference between a direct one-photon coupling and an effective two-photon coupling. The $\pi$-phase difference in the one-photon process between two enantiomers makes the interference constructive for one enantiomer but destructive for the other. Therefore only one enantiomer is excited into a higher rotational state while the other remains in the ground state. The scheme is of flexibility in the pulse waveforms and the time order of two paths. We simulate the scheme in a sample of cyclohexylmethanol~(C$_7$H$_{14}$O) molecules. Simulative results show the robust and high-fidelity ESST can be obtained when experimental concerns are considered. Finally, we propose to employ the finished ESST in implementing enantio-separation and determining enantiomeric excess.
\end{abstract}
\maketitle

\section{Introduction}
Although two enantiomers, a chiral molecule and its mirror image, may share many physical and chemical properties, living creatures consider them as different molecules because of their divergent biological activities and functions. This biological enantio-selectivity character of chiral molecules has been recognized as important for chemistry~\cite{Barrett2014,Kaushik2015,Jiang2017}, biotechnology~\cite{Joseph2012,INTLEKOFER2015304,10.1371/journal.pcbi.1007592}, and pharmaceutics~\cite{MA2016268,AMORIM2016277,SANGANYADO2017527,Ribeiro2020}. So the property of chirality is one of the most profound aspects of the world, and the enantiomer-selective tasks of chiral molecules, e.g., enantiomer separation, enantiomer purification, absolute configuration determination, and enantiomeric excess determination, are of great significance.
Except the widely used pure chemical means~\cite{Boden,McKendry,Rikken,Zepik,Bielski1,Bielski2}, optical methods are potential candidates for performing the enantiomer-selective tasks of chiral molecules~\cite{Cameron_2014,Bradshaw2015,Bradshaw:15,PhysRevA.94.032505,Bradshaw:15,PhysRevLett.122.223201,PhysRevLett.123.243202}. Some spectroscopic techniques have been established for the determination of the absolute configuration and enantiomeric excess of a chiral sample, such as circular dichroism~\cite{B515476F}, vibrational circular dichroism~\cite{Nafie2011}, and Raman optical activity~\cite{BARRON20077}. However, these techniques arise from the interference between electric-dipole and weak magnetic-dipole transitions, or in some cases involve the interference between the electric-dipole and weak electric-quadrupole transitions. Therefore these techniques require generally high-density samples~\cite{Barron2004,Nafie2013,Patterson2014}. For the enhanced probe signals, chiral molecules have been widely studied using spectroscopic techniques, e.g., ultrafast resonant X-ray spectroscopy~\cite{doi:10.1063/1.4974260,Rouxel2018} and Raman optical activity by coherent anti-Stokes Raman scattering spectral interferometry~\cite{Hiramatsu:13,Begzjav:19}.

Recently, alternative schemes based on a cyclic three-state configuration consisting of three rotational transitions in the microwave regime are considered promising. A method of coherently controlled adiabatic passage~\cite{Kral2001,Kral2003,Kral2007}, termed ``cyclic population transfer"~(CPT), is of importance and interest. Though this method may enable the highly efficient enantiomer-selective tasks by following the concepts from the adiabatic passage techniques~\cite{Bergmann1998,Vitanov2017}, while it is usually slow and complicated. Then fast schemes of nonadiabatic dynamic were proposed using resonant ultrashort pulses~\cite{Li2008,Jia_2010,PhysRevA.100.033411}. Lately, a lot of remarkable experiments were contributed to the verification of enantiomeric differentiation and the probe of enantiomeric excess based on the technique of microwave three-wave mixing~(M3WM)~\cite{Patterson2013,Patterson2013PRL,Shubert2014,Lobsiger2015,Shubert2015,Eibenberger2017,Perez2017,Perez2018}. Promoted by experimental achivements, some proposals were put forward very recently for the related issues, such as the design of cyclic three-level configurations in chiral molecules~\cite{Ye2018}, shortcut-to-adiabatic~(STA) enantiomer-selective population transfer of chiral molecules by means of shaped pulses~\cite{Vitanov2019,wu2019robust}, and dynamic methods for the enantiomeric excess determination~\cite{PhysRevA.100.033411} and the inner-state enantio-separation~\cite{PhysRevA.100.043403}.

The enantiomer-selective state transfer~(ESST) is the premise of many schemes for enantiomer separation~\cite{Vitanov2019}, enantiomer purification~\cite{Kral2001,Kral2003}, and enantiomeric excess determination~\cite{Eibenberger2017,Perez2017}.
In this work, a two-path interference approach is proposed for the ESST of chiral molecules in a cyclic three-state configuration $|1\rangle\leftrightarrow|2\rangle\leftrightarrow|3\rangle\leftrightarrow|1\rangle$ among three rotational levels. Two enantiomers are both prepared initially in the ground state $|1\rangle$, but finally evolve along two interfering paths into entirely different states. The two interfering paths are formed by a direct one-photon $|1\rangle\leftrightarrow|3\rangle$ coupling and an effective two-photon coupling with the intermediate state $|2\rangle$. It is the $\pi$-phase difference between two enantiomers in the one-photon coupling that makes the two-path interference constructive for one enantiomer but destructive for the other. Therefore only one of two enantiomers is excited into the excited state $|3\rangle$ but the other remains in the ground state $|1\rangle$.

The present scheme is performed in the nonadiabatic regime and thus faster than slow adiabatic schemes~\cite{Kral2001,Kral2003}, whilst different from the nonadiabatic dynamic methods~\cite{Li2008,Jia_2010,Eibenberger2017,Perez2017,Perez2018,PhysRevA.100.043403}, the one- and two-photon processes can be performed synchronously or subsequently, without relying on a fixed time order. Besides, by means of pulse engineering we use the shaped pulses in the scheme so as to enhance the efficiency and robustness. In addition, compared with the STA schemes~\cite{Vitanov2019,wu2019robust}, the present approach is of flexibility in the waveforms as well as the time order of two coupling paths. As an example, we apply this two-path interference scheme in a sample of cyclohexylmethanol~(C$_7$H$_{14}$O) molecules and the experimental issues concerning unwanted transitions, imperfect initial state, pulse shaping, control errors and finite lifetimes of higher energy levels are discussed. Simulative results indicate that the robust and highly efficient ESST of cyclohexylmethanol molecules can be obtained. Furthermore, based on the finished ESST, the further possible tasks of the enantiomer separation and the M3WM-assisted determination of enantiomeric excess are discussed.

\section{Two-path interference approach}\label{S2}
\subsection{Two enantiomers interacting with three orthogonal microwave fields}
\begin{figure}[htb]\centering
\centering
\includegraphics[width=\linewidth]{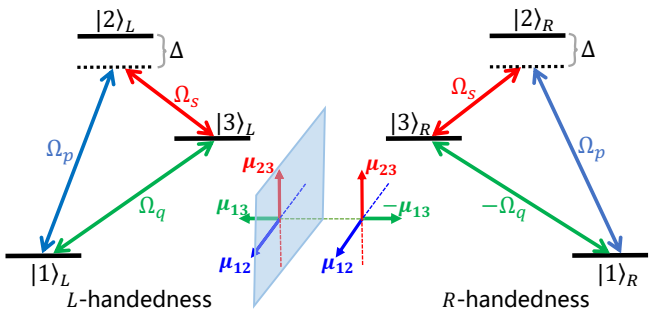}
\caption{Couplings among three rotational energy states in two enantiomers of chiral molecules. Three orthogonal microwave fields are imposed to drive three types of transitions with orthogonal electric-dipole moments, respectively. The electric-dipole moments of two enantiomers are mirrored with each other.}\label{f1}
\end{figure}
A cyclic three-state configuration in two enantiomers is shown in Fig.~\ref{f1}, in which three rotational energy states $|1\rangle$, $|2\rangle$ and $|3\rangle$ are considered for each enantiomer. The electric-dipole-allowed transitions $|1\rangle\leftrightarrow|2\rangle$, $|1\rangle\leftrightarrow|3\rangle$, and $|2\rangle\leftrightarrow|3\rangle$ have their individual dipole moments ${\overrightarrow\mu}_{12}$, ${\overrightarrow\mu}_{13}$, and ${\overrightarrow\mu}_{23}$. Because the inertia moments of molecules determine their rotational spectra, molecular rotational spectra depend on the distribution of atomic masses in the molecules. Therefore, the rotational spectroscopy is sensitive to even tiny structural or mass changes. Correspondingly, three types of electric-dipole moments of one enantiomer are mirrored with their counterparts of the other enantiomer due to their asymmetric distribution of atoms~\cite{2012PJA8803B-06}.

Three orthogonal microwave fields $P$, $Q$, and $S$ irradiate molecules of the two enantiomers to drive the three rotational transitions with individual Rabi frequencies $\Omega_p$, $\Omega_q$, and $\Omega_s$, as shown in Fig.~\ref{f1}. The strength of the $j$ field~($j=p,q,s$) is expressed by $\overrightarrow{E_j}= \overrightarrow{e_j}{\varepsilon}_j\cos(\omega_jt+\phi_j)$, where $\overrightarrow{e_j}$, ${\varepsilon}_j$, $\omega_j$, and $\phi_j$ are the unit vector, amplitude, frequency, and phase, respectively. Correspondingly, $\Omega_p\equiv{\overrightarrow\mu}_{12}\cdot\overrightarrow{e_p}{\varepsilon}_p$, $\Omega_q\equiv{\overrightarrow\mu}_{13}\cdot\overrightarrow{e_p}{\varepsilon}_p$, and $\Omega_s\equiv{\overrightarrow\mu}_{23}\cdot\overrightarrow{e_p}{\varepsilon}_p$.

Here we set $|1\rangle$ as the zero-energy point and use the natural unit $\hbar=1$ for simplicity.
The Hamiltonian of the cyclic three-state configuration can be represented by
\begin{eqnarray}\label{e1}
\hat H_0&=&\omega_{12}|2\rangle\langle 2|+\omega_{13}|3\rangle\langle 3|+\Big({\overrightarrow\mu}_{12}\cdot\overrightarrow{E_p}|1\rangle\langle 2|\nonumber\\&&+{\overrightarrow\mu}_{13}\cdot\overrightarrow{E_q}|1\rangle\langle 3|+{\overrightarrow\mu}_{23}\cdot\overrightarrow{E_s}|2\rangle\langle 3|+{\rm H.c.}\Big),
\end{eqnarray}
where $\omega_{1,n}~(n=2,3)$ is the $|1\rangle\leftrightarrow|n\rangle$ transition frequency. Owing to the mirror reflection of electric-dipole  moments of two enantiomers, the triple product ${\overrightarrow\mu}_{12}\cdot({\overrightarrow\mu}_{13}\times{\overrightarrow\mu}_{23})$ is of opposite signs for the two enantiomers. For convenience, we specify the model such that for two enantiomers, ${\overrightarrow\mu}_{12}$ as well as ${\overrightarrow\mu}_{23}$ is of an identical orientation while ${\overrightarrow\mu}_{13}$ is of opposite orientations, as shown in Fig.~\ref{f1}, which results in two enantiomers holding identical $\Omega_p$ and $\Omega_s$ but opposite-sign $\Omega_q$. In laboratory framework, it can be achieved by choosing a suitable set of axes, as treated in the recent experiment~\cite{Eibenberger2017}. Alternatively, one can choose a suitable group of phases of the three microwave fields, as the recent experiment~\cite{Perez2018}.
Hereinafter we consider conditions: (i)~Rotating-wave approximation $\omega_{j}\gg|\Omega_j|$; (ii)~Resonant one-photon $|1\rangle\leftrightarrow|3\rangle$ transition $\omega_{13}=\omega_{q}$; (iii)~Resonant two-photon $|1\rangle\leftrightarrow|3\rangle$ transition but off-resonant one-photon $|1\rangle\leftrightarrow|2\rangle$ and $|2\rangle\leftrightarrow|3\rangle$ transitions $\omega_{12}-\omega_{p}=\omega_{23}-\omega_{s}=\Delta$.

\subsection{Enantiomer-selective state transfer through the two-path interference}
Under the rotating-wave approximation, the Hamiltonian~(\ref{e1}) can be written in the interaction picture for two enantiomers
\begin{equation}\label{e2}
\hat H_{L,R}=\left(\frac{\Omega_p}2|2\rangle\langle 1|+\frac{\Omega_s}2|2\rangle\langle 3|\right)e^{i\Delta t}\pm\frac{\Omega_q}2|1\rangle\langle 3|+{\rm H.c.},
\end{equation}
where the subscripts ``$L$" and ``$R$" denote the left and right handedness, respectively. We have chosen all phases of three microwave pulses as zero. The last term describes a direct $|1\rangle\leftrightarrow|3\rangle$ coupling of a one-photon process, and two enantiomers differ in the Rabi frequency $\Omega_q$ by a $\pi$ phase. The identical first term denotes the detuned $|1\rangle\leftrightarrow|2\rangle$ and $|2\rangle\leftrightarrow|3\rangle$ transitions with detuning $\Delta$, but the two-photon $|1\rangle\leftrightarrow|3\rangle$ transition is resonant.

We aim to form an effective two-photon $|1\rangle\leftrightarrow|3\rangle$ coupling that can interfere with the direct one-photon coupling. Therefore, the interference between the one-photon transitions $|1\rangle\leftrightarrow|3\rangle$ and $|1\rangle\leftrightarrow|2\rangle$ or $|2\rangle\leftrightarrow|3\rangle$ has to be absent, because it makes the one-photon transition $|1\rangle\leftrightarrow|3\rangle$ not only governed by $Q$ pulse but also affected by $P$ and $S$ pulses. A valid way is to eliminate the occupation of the intermediate state $|2\rangle$ during the two-photon process $|1\rangle\leftrightarrow|2\rangle\leftrightarrow|3\rangle$, which can make the one- and two-photon processes independent on each other. Then under the large-detuning condition $\Delta\gg\Omega_p,\Omega_s$ the dynamics of two enantiomers can be dominated by the following effective Hamiltonian~(see {\bf Appendix A} for details)
\begin{equation}\label{e3}
\hat H_{L,R}^e=\frac{\Omega_q\mp\Omega_{\rm eff}}2|1\rangle\langle 3|+{\rm H.c.},
\end{equation}
with $\Omega_{\rm eff}\equiv{\Omega_p\Omega_s}/{2\Delta}$. Since two enantiomers differ in the one-photon coupling by a $\pi$-phase difference, the two paths interfere with each other constructively for $R$-handed molecules but destructively for $L$-handed molecules. We assume that the operation starts at $t=0$ and ends at $t=T$. If we control the pulse areas of the three microwave pulses to satisfy
\begin{equation}\label{e4}
\int_0^T\frac{\Omega_q-\Omega_{\rm eff}}2dt=n_l\pi,~\int_0^T\frac{\Omega_q+\Omega_{\rm eff}}2dt=(n_r+\frac12)\pi,
\end{equation}
with $n_l$ and $n_r$ being integer numbers, $R$-handed molecules will be excited into $|3\rangle$ while $L$-handed molecules remain or evolve back to $|1\rangle$, which means the implementation of the ESST. This ESST is achieved by the interference of transitions rather than the CPT commonly used in the existing schemes~\cite{Kral2001,Kral2003,Li2008,Jia_2010,Perez2017,Perez2018,PhysRevA.100.043403,Vitanov2019,wu2019robust}.

\section{Pulse engineering}
\subsection{Efficiency of the state transfer with shaped pulses}
High efficiency is one of the advantages of the enantiomer-selective schemes based on the microwave-regime rotational spectroscopy, because the pure electric-dipole couplings allow for the strong molecule-filed interactions and the intense chiral molecular signals. In order to ensure a faithful two-path interference, the ratio $\Delta/\Omega_0$ in theory is supposed to be as large as possible, with $\Omega_0\equiv\max\{\Omega_p,~\Omega_s\}$ being the maximum amplitude of Rabi frequencies $\Omega_p$ and $\Omega_s$. In practice, however, it is unnecessary to adopt a very large ratio $\Delta/\Omega_0$, and thus a relatively high efficiency can be held.

\begin{figure}\centering
\includegraphics[width=\linewidth]{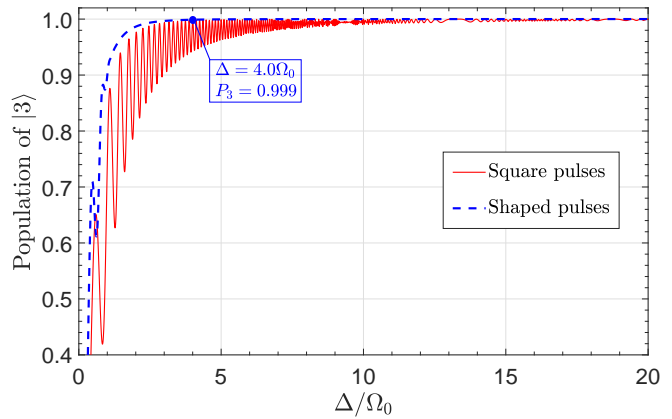}
\caption{Final population of $|3\rangle$ of all molecules excited through only the two-photon process with varying $\Delta/\Omega_0$ for square~(solid red line) or shaped~(blue dashed line) $P$ and $S$ pulses.}\label{f2}
\end{figure}
In order to get a moderate ratio $\Delta/\Omega_0$, now we solely consider the two-photon coupling to excite all molecules from $|1\rangle$ into $|3\rangle$, and the partial Hamiltonian $\hat h(t)=(\Omega_p|2\rangle\langle 1|+\Omega_s|2\rangle\langle 3|)e^{i\Delta t}/2+{\rm H.c.}$ governs the evolution. For the square $P$ and $S$ pulses with $\Omega_p=\Omega_s=\Omega_0$, the pulse area should satisfy $\int_0^T\Omega_{\rm eff}dt=\pi$, i.e., $T=2\pi\Delta/\Omega_0^2$, to make all molecules excited into $|3\rangle$. In Fig.~\ref{f2}~(solid red line), we plot the final~($t=T$) population of $|3\rangle$ of all molecules with varying $\Delta/\Omega_0$.
When the ratio $\Delta/\Omega_0$ reaches over a certain value~(about 10), the final population of $|3\rangle$ reaches near unity. For $\Delta=10\Omega_0$, the effective strength of the two-photon $|1\rangle\leftrightarrow|3\rangle$ coupling is $\Omega_{\rm eff}=0.05\Omega_0$, which is smaller than the original strength by near two orders of magnitude.

Recently, pulse engineering is widely studied to seek for robust dynamics and steady quantum state~\cite{Glaser2015,PhysRevLett.123.100501,Wu_2019}. Here,  we use the pulse engineering to enhance the efficiency of the ESST by replacing square pulses with shaped pulses that are turned on and off smoothly. The waveforms of $P$ and $S$ pulses can be chosen as a single-period $\cos$-like function
\begin{eqnarray}\label{e5}
\Omega_p=\Omega_s=\left\{\begin{array}{cc}
\frac{\Omega_0}2\left(1-\cos\frac{2\pi t}{T_0}\right),&0\leqslant t\leqslant T_0\\
0,&{\rm otherwise}
\end{array}
\right.,
\end{eqnarray}
with $T_0$ being the period. Then with $T=T_0=16\pi\Delta/3\Omega_0^2$, the final population of $|3\rangle$ with varying $\Delta/\Omega_0$ is also plotted in Fig.~\ref{f2}~(dashed blue line). Apparently, the final population of $|3\rangle$ can reach very near unity as long as $\Delta/\Omega_0>2$. There are few oscillations once the final population of $|3\rangle$ reach near unity, which denotes a robust state transfer. Furthermore, when $\Delta/\Omega_0>4$ the final population of $|3\rangle$ can be over 0.999. For $\Delta=4\Omega_0$, the two-photon coupling strength is $\Omega_{\rm eff}=0.125\Omega_0$ smaller than the original strength by just one order of magnitude.

The strengths of magnetic-dipole and electric-quadrupole transitions are five and six orders of magnitude, weaker than that of the electric-dipole transition, respectively. Therefore, the efficiency of the present scheme based on the square pulses is about three or four orders of magnitude higher than that of the conventional circular dichroism, and comparable to the photoelectron circular dichroism~\cite{doi:10.1002/9780470259474.ch5}. The scheme with pulse engineering can work better than the photoelectron circular dichroism, and the efficiency is over four orders of magnitude higher than that of the conventional circular dichroism.

\subsection{Flexibility in the pulse waveforms and the time order of two paths}
With the established two interfering paths, the conditions of achieving the ESST are listed in Eq.~(\ref{e4}). The conditions are not limited by the waveforms of three pulses and even the time order of two paths, which is different from the existing adiabatic~\cite{Kral2001,Kral2003}, nonadiabatic~\cite{Li2008,Eibenberger2017,Perez2017,Perez2018,PhysRevA.100.043403}, and STA schemes~\cite{Vitanov2019,wu2019robust}. To test it, we choose different waveforms of $Q$ pulse combined with the the waveforms in Eq.~(\ref{e5}) of $P$ and $S$ pulses to exhibit the flexibility in the pulse waveforms and the time order of two paths.  Typically, we pick up $n_l=n_r=0$ hereinafter so as to achieve the state transfer in a short duration, and it requires $\int_0^T\Omega_qdt=\int_0^T\Omega_{\rm eff}dt={\pi}/2$
which gives $T_0=8\pi\Delta/3\Omega_0^2$. Take several examples of the waveform of $Q$ pulse:

(i)~Single-period $\cos$-like waveform imposed after finishing the two-photon effective coupling
\begin{eqnarray}\label{e6}
\Omega_q=\left\{\begin{array}{cc}
\frac{\Omega'_0}2\left[1-\cos\frac{2\pi (t-T_0)}{T'_0}\right],&T_0\leqslant t\leqslant T_0+T'_0\\
0,&{\rm otherwise}
\end{array}
\right.
\end{eqnarray}
with $\Omega'_0$ being the maximum amplitude, $T_0$~(the period of $\Omega_p$ and $\Omega_s$) the delay, and $T'_0=\pi/\Omega'_0$ the period, making $T=T_0+T'_0$;

(ii)~Gaussian waveform with a delay to $\Omega_{\rm eff}$
\begin{eqnarray}\label{e7}
\Omega_q=\Omega'_0\exp\left[-(t-3t_c)^2/t_c^2\right],
\end{eqnarray}
with $t_c=\sqrt\pi/2\Omega'_0$ being the width, $T=\max\{6t_c, T_0\}$;

(iii)~Single-period $\cos^2$-like waveform thoroughly coinciding with $\Omega_{\rm eff}$
\begin{eqnarray}\label{e8}
\Omega_q=\left\{\begin{array}{cc}
\frac{\Omega'_0}4\left(1-\cos\frac{2\pi t}{T_0}\right)^2,&0\leqslant t\leqslant T_0\\
0,&{\rm otherwise}
\end{array}
\right.
\end{eqnarray}
with $\Omega'_0=\Omega_0^2/2\Delta$ being the maximum amplitude, $T=T_0$.

\begin{figure}\centering
\includegraphics[width=\linewidth]{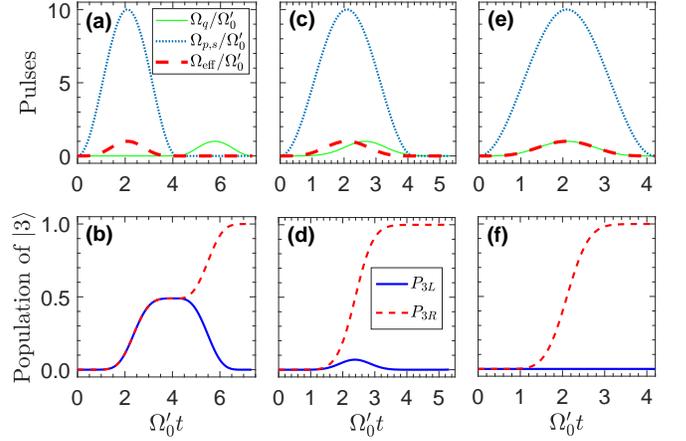}
\caption{(a), (c), and (e): Shapes of $\Omega_p$, $\Omega_s$, $\Omega_q$ and $\Omega_{\rm eff}$. For $\Omega_q$, the shapes in (a), (c), and (e) are of the waveforms in Eqs.~(\ref{e6}), (\ref{e7}), and (\ref{e8}), respectively. (b), (d), and (f): Time evolutions of the population of $|3\rangle$ for the left-handed~(solid blue line) and right-handed~(dashed red line) molecules. Parameters: $\Omega_0=10\Omega'_0$ and $\Delta=50\Omega'_0$.}\label{f3}
\end{figure}
Three exampled waveforms in Eqs.~(\ref{e6})-(\ref{e8}) of $Q$ pulse combined with $\Omega_{\rm eff}$ and the waveform in Eq.~(\ref{e5}) of $P$ and $S$ pulses are shown, respectively, in Figs.~\ref{f3}(a), (c), and (e), by means of which we plot the time evolutions of the population of $|3\rangle$ for two enantiomers~($P_{3L}$ for $L$-handedness and $P_{3R}$ for $R$-handedness) in Figs.~\ref{f3}(b), (d), and (f), correspondingly. We learn that all the three sets of pulse waveforms can guarantee that $R$-handed molecules are excited into $|3\rangle$ from $|1\rangle$ but $L$-handed ones unchanged. It illustrates that the present two-path interference scheme of the ESST possesses flexibility in the pulse waveforms and the time order of two paths.

\section{Simulation with cyclohexylmethanol molecules}\label{S4}
\begin{figure}[htb]\centering
\includegraphics[width=\linewidth]{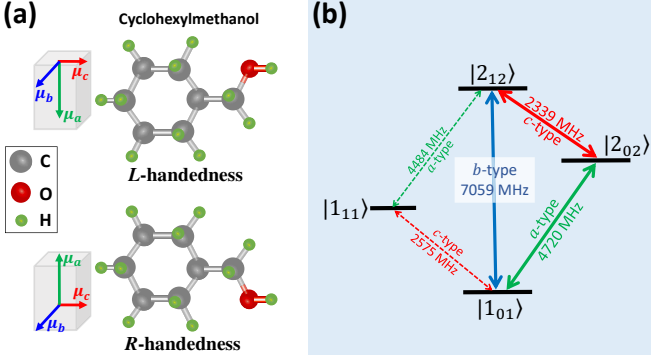}
\caption{(a)~Mirror-refection molecular structure diagram of two enantiomers of cyclohexylmethanol molecules. The electric-dipole moment orientations of three types of rotational transitions are depicted in inner cubes nearby. (b)~Cyclic three-state configuration~(solid thick arrows) in cyclohexylmethanol molecules and unwanted transitions~(dashed thin arrows).}\label{f4}
\end{figure}

In this section, we simulate the two-path interference scheme by using a cyclic three-state configuration $|1_{01}\rangle\leftrightarrow|2_{12}\rangle\leftrightarrow|2_{02}\rangle\leftrightarrow|1_{01}\rangle$ in cyclohexylmethanol~(C$_7$H$_{14}$O) molecules, which has been used in a supersonic jet experiment for the CPT within 2-8~GHz microwave regime~\cite{Perez2018}. The mirror-refection molecular structure diagram of two enantiomers is shown in Fig.~\ref{f4}(a), and the cyclic three-state configuration in Fig.~\ref{f4}(b). The energy levels designated with $|J_{K_a,K_c}\rangle$, where $J$ is the rotational quantum number and $K_a$ and $K_c$ are the projections of $J$ onto the principal axes of the molecule. Two enantiomers have the conformer constants $A=3898.45$~MHz, $B=1319.59$~MHz, and $C=1062.55$~MHz. There are three types of rotational transitions, $a$-type, $b$-type, and $c$-type with dipole moments $|\mu_a|=0.4$~Debye, $|\mu_b|=1.2$~Debye, and $|\mu_c|=0.8$~Debye, respectively. We define $|1\rangle\equiv|1_{01}\rangle$, $|2\rangle\equiv|2_{12}\rangle$ and $|3\rangle\equiv|2_{02}\rangle$ to make Figs.~\ref{f1} and \ref{f4}(a) coincident. Another state $|4\rangle\equiv|1_{11}\rangle$ is also considered, because it is involved in two possible unwanted transitions. In the following, by means of the master equation~(see {\bf Appendix B} for details) we simulate the effect of some possible factors on the performance of the two-path interference scheme of the ESST.

\subsection{Pulse shaping with a finite time resolution}
\begin{figure}\centering
\includegraphics[width=\linewidth]{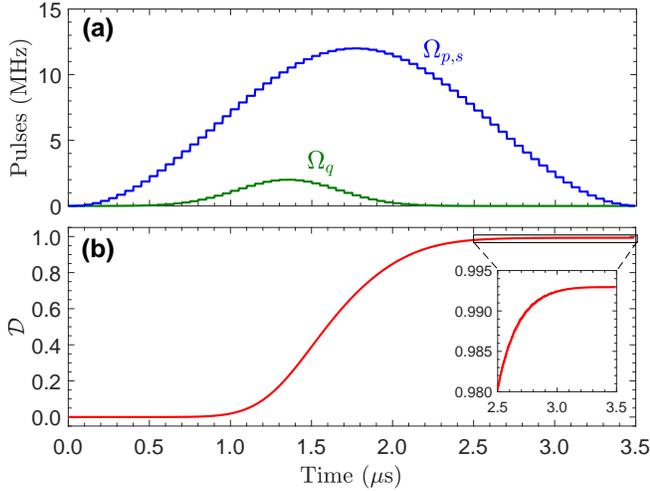}
\caption{(a)~Waveforms of three pulses with a time resolution $dt=50$~ns. (b)~Time evolution of $\mathcal{D}$ by using the pulse waveforms in (a). Parameters without energy relaxations: $\Omega_0=12$~MHz, $\Omega'_0=2$~MHz, $T_0=3.5~\mu$s, $t_c=0.443~\mu$s, and $\Delta=60$~MHz.}\label{f5}
\end{figure}

With arbitrary waveform generators, the shapes of three pulses can be obtained by modulating the filed amplitudes corresponding to voltages applied to the electro-optic modulators~\cite{PhysRevLett.110.240501,BBZhou}. Referring to three transition dipole moments, we adopt $\Omega_0=12$~MHz and $\Omega'_0=2$~MHz to simulate the ESST of cyclohexylmethanol molecules. The waveforms of three pulses given in Eqs.~(\ref{e5}) and (\ref{e7}) are adopted, with $T=T_0=3.5~\mu$s and $t_c=0.443~\mu$s.

Continuously varied waveforms are desired but usually unrealistic. In Fig.~\ref{f5}(a), we plot the waveforms of three pulses by introducing a finite time resolution $dt=50$~ns, and thus each waveform consists of a series of square pulses within $50$~ns duration. Then we substitute the waveforms in Fig.~\ref{f5}(a) into the master equation and consider the energy relaxations absent. The time evolution of the ESST of the cyclohexylmethanol sample is exhibited in Fig.~\ref{f5}(b), for which the performance~(or fidelity) of the ESST is measured by a quantity
\begin{eqnarray}\label{e9}
\mathcal{D}=\left|P_{3L}-P_{3R}\right|.
\end{eqnarray}
The fidelity of the ESST increases with time, and a high-fidelity ESST with $\mathcal{D}=0.993$ is indicated finally. As a matter of fact, the time resolution $dt=50$~ns used here is far large than that reported in recent experiments. The arbitrary waveform generators can realize the minimal possible time resolution $\sim0.25$~ns~\cite{PhysRevLett.110.240501} or even $\sim0.1$~ns~\cite{BBZhou}. Therefore in the following, we use safely the time resolution $dt=10$~ns for the waveforms of three pulses.

\subsection{Errors in amplitudes and frequencies of pulses}
In experiment, it is inevitable that operations on three pulses suffer from errors originating from the imprecise apparatus, imperfect control, and various unpredictable fluctuations, so it is essential to investigate the effect of errors in three pulses on the ESST performance. Here, we mainly concern random amplitude noises and frequency drifts.

\begin{figure}\centering
\includegraphics[width=\linewidth]{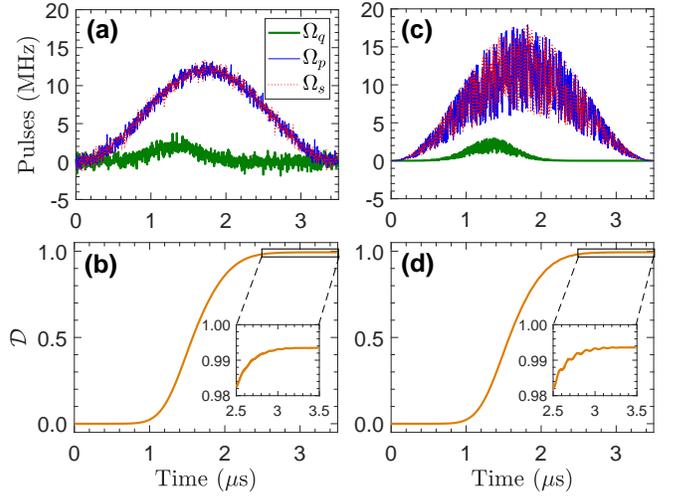}
\caption{(a)~Waveforms of three AWGN-mixed pulses with $R_{\rm SN}=10$. (b)~Time evolution of $\mathcal{D}$ by using the pulse waveforms in (a). (c)~Waveforms of three randomly-fluctuated pulses with $\eta=0.5$. (d)~Time evolution of $\mathcal{D}$ by using the pulse waveforms in (c). $dt=10$ ns and other parameters are the same as Fig.~\ref{f5}.}\label{f6}
\end{figure}
Due to some uncontrollable factors such as the stray light mixing and unstability of voltages, the three pulses may be disturbed by additive white Gaussian noises~(AWGN) and random fluctuations.
An AWGN-mixed Rabi frequency is written as
\begin{eqnarray}\label{e10}
\Omega_{\rm AWGN}(t)&=&\Omega(t)+{\rm awgn}[\Omega(t),~R_{\rm SN}],
\end{eqnarray}
with ${\rm awgn}$ being the generation function of AWGN mixed into the original pulse $\Omega(t)$ with a signal-to-noise ratio $R_{\rm SN}$.
A randomly-fluctuated Rabi frequency is
\begin{eqnarray}\label{e11}
\Omega_{\rm rand}(t)&=&\Omega(t)[1+{\rm rand}(t,~\eta)],
\end{eqnarray}
where ${\rm rand}$ is a function generating a random number within $[-\eta,~\eta]$.
We plot three AWGN-mixed waveforms with $R_{\rm SN}=10$ and randomly-fluctuated waveforms with $\eta=0.5$, respectively, in Figs.~\ref{f6}(a) and (c), based on which the time evolutions of the ESST performance of the cyclohexylmethanol sample are plotted in Figs.~\ref{f6}(b) and (d), correspondingly. Apparently, the waveforms in Figs.~\ref{f6}(a) and (c) are deformed greatly, but the time evolutions of the ESST performance is influenced little by either the AWGN or random fluctuations in three pulses. And the high-fidelity ($\mathcal{D}=0.994$) ESST can be reached finally. In fact, the realistic signal-to-noise ratio is usually lower than that considered in Fig.~\ref{f6}. Figure~\ref{f6} demonstrates that the influence of the random amplitude noises including AWGN and random fluctuations on the ESST performance are negligible. The reason lies in that AWGN as well as random fluctuations has random absolute values and random plus-minus signs, which causes that the valid pulse areas for the two interfering paths remain unchanged.

\begin{figure}\centering
\includegraphics[width=\linewidth]{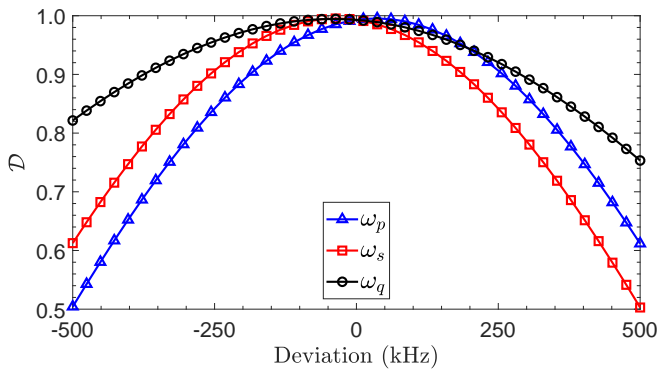}
\caption{Effect of the deviation of each pulse frequency on the ESST performance of the cyclohexylmethanol sample. Parameters are the same as Fig.~\ref{f6}.}\label{f7}
\end{figure}
\begin{figure}[htb]\centering
\includegraphics[width=\linewidth]{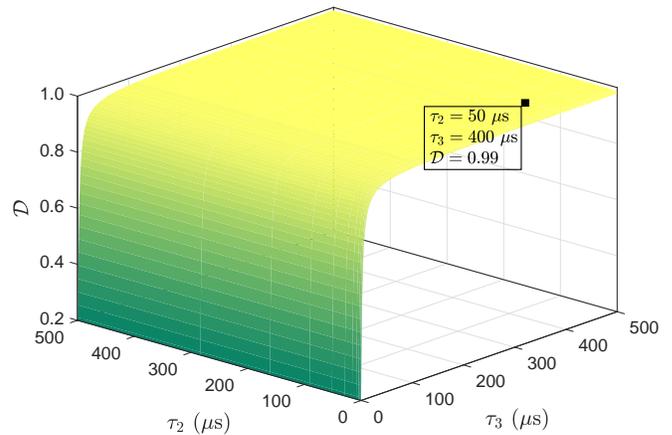}
\caption{Final ESST performance of the cyclohexylmethanol sample with varying $\tau_2$ and $\tau_3$. Parameters are the same as Fig.~\ref{f6}.}\label{f8}
\end{figure}
\begin{figure*}\centering
\includegraphics[width=0.9\linewidth]{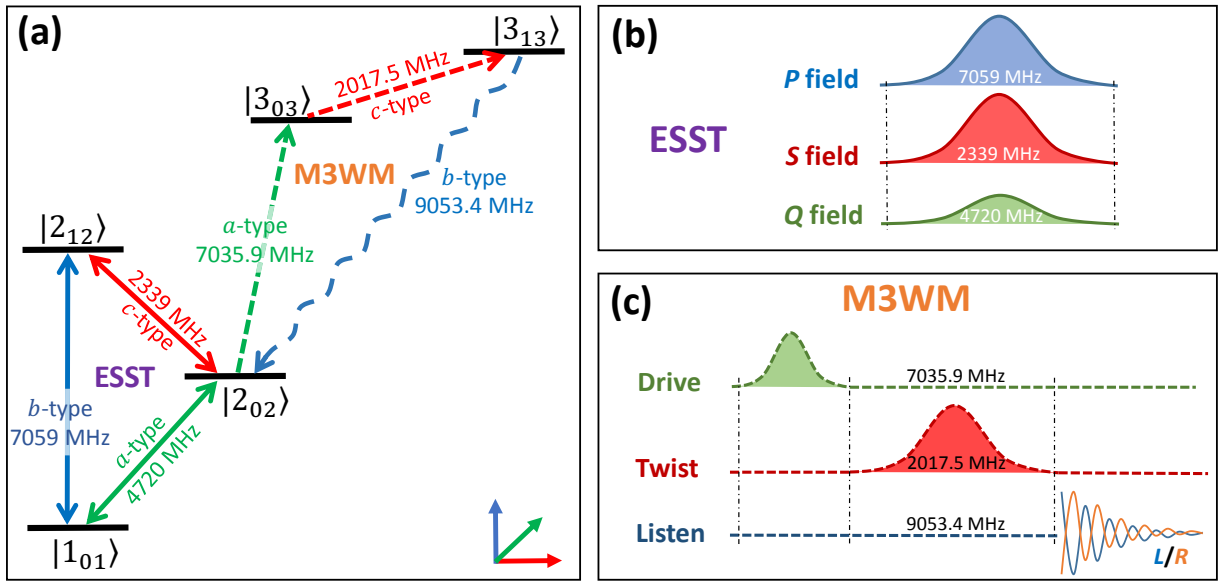}
\caption{(a)~Two three-state cyclic configurations of combining the ESST and the M3WM. (b)~Diagram of the pulse scheme for the ESST. (c)~Diagram of the pulse scheme for the M3WM.}\label{f9}
\end{figure*}
Due to the imprecise apparatus or imperfect control of generating three pulses at their fixed frequencies, the pulse frequencies may drift to some extent.
The effect of the deviation of each pulse frequency on the ESST performance of the cyclohexylmethanol sample is plotted in Fig.~\ref{f7}, from which we learn that the ESST performance is sensitive to the frequency deviation of each pulse~(i.e., detuning of the corresponding transition) since the two-path interference approach is built on the resonant regime for both one-photon and two-photon coupling paths. A larger ratio of detuning to Rabi frequency will usually spoil the intended dynamics more significantly. In Fig.~\ref{f7}, however the line for the frequency deviation of $Q$ pulse that has the smallest Rabi frequency holds the highest $\mathcal{D}$, which is because that the effective Rabi frequency~($\max\{\Omega_{\rm eff}\}=1.2$ MHz) of the two-photon coupling path is less than $\Omega_q$. To summary, the present approach requires relatively precise control of pulse frequencies. As a matter of fact, it is a common issue for the existing schemes~\cite{Kral2001,Kral2003,Li2008,Eibenberger2017,Perez2017,Perez2018,PhysRevA.100.043403,Vitanov2019,wu2019robust}, especially for the schemes in resonant regime~\cite{Li2008,Eibenberger2017,Perez2017,Perez2018,Vitanov2019,wu2019robust}.
To this end, the state-of-the-art quantum optimal control techniques may give some alternative insights~\cite{Glaser2015,PhysRevLett.123.100501}.

\subsection{Energy relaxation}
The states of cyclohexylmethanol molecules used for the cyclic three-state configuration are three relatively lower rotational levels and have long coherence times. However, there is a little probability that higher states relax into the lower states than them. Here we consider the lifetimes of $|2\rangle$ and $|3\rangle$ as $\tau_2$ and $\tau_3$, respectively, and then $\gamma_{1,2}=|\mu_b|/[\tau2(|\mu_a|+|\mu_b|+|\mu_c|)]$, $\gamma_{3,2}=|\mu_c|/[\tau2(|\mu_a|+|\mu_b|+|\mu_c|)]$, $\gamma_{4,2}=|\mu_a|/[\tau2(|\mu_a|+|\mu_b|+|\mu_c|)]$ and $\gamma_{1,3}=1/\tau_3$. For simplicity, the rotational state $|4\rangle$ is assumed to be steady since  $|4\rangle$ is a lower state than $|2\rangle$ and $|3\rangle$, and besides $|4\rangle$ is populated very little.

Figure~\ref{f8} shows the effect of varying $\tau_2$ and $\tau_3$ on the ESST performance of the cyclohexylmethanol sample.
Apparently, even a near zero $\tau_2$ hardly affects the performance of the ESST since $|2\rangle$ is not populated during the whole enantiomer-selective process. For achieving the high-fidelity~($\mathcal{D}>0.99$) ESST of the cyclohexylmethanol sample, $\tau_2>50~\mu s$ and $\tau_3>400~\mu s$ are available enough. The lifetimes of the rotational states are seldom considered one of the the main factors that affect the performance of the enantiomer-selective tasks of chiral molecules for the schemes based on the microwave spectroscopy. On one hand, the electric-dipole couplings allow the fast implementation of the enantiomer-selective tasks. On the other hand, rotational states that are adopted to form a cyclic three-state configuration are of long radiative lifetimes which is usually over 1~microsecond.

\section{Further applications of the finished ESST}
The ESST finished through the two-path interference scheme renders the discrimination of chiral molecules, which can be then followed by other enantiomer-selective tasks of chiral molecules. Not limited by the enantiomeric excess of the sample, the molecules of a certain handedness~(either $L$- or $R$-handedness is possible, see {\bf Appendix C}) can be selectively excited into the higher state $|3\rangle$, while the molecules of the other handedness remain in the lower state $|1\rangle$. Owing to the energy difference between two enantiomers, the molecules excited into the higher state $|3\rangle$ can be selectively ionized by applying a resonantly enhanced multi-photon ionization scheme, and then the enantio-separation can be implemented by an electric field to remove the ionized molecules~\cite{Vitanov2019,PhysRevA.77.030702}.

In addition, the finished ESST can be applied to determine the enantiomeric excess by combining the technique of M3WM. Taking the cyclohexylmethanol molecules as an example, an extra three-state cyclic configuration $|2_{02}\rangle\leftrightarrow|3_{03}\rangle\leftrightarrow|3_{13}\rangle\leftrightarrow|2_{02}\rangle$ are introduced to serve for the M3WM, as shown in Fig.~\ref{f9}(a). First of all, the ESST is executed with synchronous three pulses in the two-path interference scheme, as shown in Fig.~\ref{f9}(b).
Then two new microwave fields should be imposed for the M3WM. The earlier one at the frequency $7035.9$~MHz serves as the so-called ``drive" pulse with a pulse area $\pi/2$, exciting the molecules in $|2_{02}\rangle$ to make them evolve into the maximum coherence between $|2_{02}\rangle$ and $|3_{03}\rangle$. A following ``twist" pulse at $2017.5$~MHz with a pulse area $\pi$ shifts the population from $|3_{03}\rangle$ into $|3_{13}\rangle$ so as to create a coherence between the states $|2_{02}\rangle$ and $|3_{13}\rangle$.

Afterward, in the form of the free-induction decay of the sample, the molecular signal that is mutually orthogonal to the drive and twist fields is created and can be recorded using Fourier transform microwave spectroscopy techniques~\cite{doi:10.1063/1.1136443}. The diagram of the pulse scheme for the M3WM is shown in Fig.~\ref{f9}(c).
The intensity of this induced signal, i.e., so-called ``listen" field, is proportional to~\cite{doi:10.1002/anie.201307159}
\begin{equation}\label{e12}
ee\cdot|\mu_a\cdot(\mu_b\times\mu_c)|\cdot\cos\left[2\pi \omega t+\frac{\pi\mu_a\cdot(\mu_b\times\mu_c)}{2|\mu_a\cdot(\mu_b\times\mu_c)|}\right],
\end{equation}
where $ee$ denotes the enantiomeric excess, and $\omega$ the frequency of the listen transition, i.e., 9053.4~MHz. Because the sign of ${\mu_a\cdot(\mu_b\times\mu_c)}/{|\mu_a\cdot(\mu_b\times\mu_c)|}$ is enantiomer-dependent, the sign of the listen signal is dependent on the choice of the enantiomer excited in the ESST process. Finally, taking a sample with known enantiomeric excess as a reference and normalizing the signal intensity of this sample, the enantiomeric excess of a target sample can be determined according to the listen field sign and the intensity comparison of signals between the target and reference samples. It is noted that the M3WM is based on a finished ESST, so it works only for the molecules of one enantiomer that has been selectively excited in the ESST process. Therefore, even when the molecular sample is the racemic mixture of enantiomers, the definite-intensity listen signal can be detected, which is different from the conventional M3WM technique where a nonzero net signal arises only if one of the enantiomers is in excess~\cite{Patterson2013,Patterson2013PRL,Shubert2014}.

In addition, except the commonly used means with the drive pulse followed by the twist pulse, here we propose two alternative ways to the implementation of the maximum coherence for the M3WM between $|2_{02}\rangle$ and $|3_{13}\rangle$. The two ways are both based on the two-photon processes $|2_{02}\rangle\leftrightarrow|3_{03}\rangle\leftrightarrow|3_{13}\rangle$ with two synchronous pulses, one for the transition $|2_{02}\rangle\leftrightarrow|3_{03}\rangle$ with Rabi frequency $\Omega_{\rm d}$, and the other for $|3_{03}\rangle\leftrightarrow|3_{13}\rangle$ with Rabi frequency $\Omega_{\rm t}$. One way to the maximum coherence between $|2_{02}\rangle$ and $|3_{13}\rangle$ is to form an effective $|2_{02}\rangle\leftrightarrow|3_{13}\rangle$ coupling through the adiabatic elimination of $|3_{03}\rangle$ with a large one-photon detuning $\Delta'$,  which is similar to the two-photon path in the two-path interference of the ESST. Then by controlling the pulse pulse area $\int\Omega_{\rm d}\Omega_{\rm t}/2\Delta'dt=\pi/2$, the maximum coherence between $|2_{02}\rangle$ and $|3_{13}\rangle$ can be attained. The other way is based on a resonant stimulated Raman process, and the conditions are the relation $\Omega_{\rm d}=(\sqrt2+1)\Omega_{\rm t}$ and the pulse area $\int\sqrt{\Omega_{\rm d}^2+\Omega_{\rm t}^2}dt=\pi$, which consumes shorter time than the way of the second-order effective $|2_{02}\rangle\leftrightarrow|3_{13}\rangle$ coupling. Different from the commonly used means, the latter two ways implement the maximum coherence between $|2_{02}\rangle$ and $|3_{13}\rangle$ in just one step, and the intermediate coherence between $|2_{02}\rangle$ and $|3_{03}\rangle$ is unnecessary.

\section{Conclusion}\label{S5}
We present a two-path interference scheme for the enantiomer-selective state transfer~(ESST) of chiral molecules. Two interfering paths are constructed by the one- and two-photon $|1\rangle\leftrightarrow|3\rangle$ transitions. The $\pi$-phase difference in the one-photon path between two enantiomers enables the interference destructive for one enantiomer but constructive for the other, which therefore results in the entirely different population distributions of two enantiomers. The simulative application of the present scheme in a cyclohexylmethanol sample demonstrates that the high-fidelity ESST can be achieved, even if the unwanted transitions, imperfect initial state, finite time resolution in pulse shaping, control errors and finite lifetimes of higher energy levels are taken into account. Furthermore, we seek for the application of the finished ESST in implementing enantio-separation and determining enantiomeric excess.

The two-path interference scheme is distinct from the existing schemes of the enantiomer-selective state transfer of chiral molecules. On one hand, the two-path interference scheme is faster than the adiabatic schemes that need slowly varied drive fields. On the other hand, the scheme can be performed with separate, partially overlapping, or synchronous two paths, which is of more flexibility of pulse sequences than the cyclic-population-transfer schemes depending on a fixed pulse order. In addition, compared with the lately developed shortcut-to-adiabatic schemes with synchronous pulses, the two-path interference scheme is not limited by the pulse waveforms and the match relation among the pulse amplitudes. Finally, the two alternative ways to the implementation of the maximum coherence for the M3WM are different from the commonly used means.

\section*{ACKNOWLEDGEMENTS}
This work was supported by National Natural Science Foundation of China (NSFC) (11675046, 21973023, 11804308), Program for Innovation
Research of Science in Harbin Institute of Technology (A201412),
Postdoctoral Scientific Research Developmental Fund of Heilongjiang Province (LBH-Q15060), and National Basic Research Program of China (2014CB340203).

\section*{Appendix A: Effective Hamiltonian for two enantiomers}
By solving the Schr\"{o}dinger equation $i\partial|\psi(t)\rangle/\partial t=\hat H_{L,R}|\psi(t)\rangle$, where $|\psi(t)\rangle=c_1|1\rangle+c_2|2\rangle+c_3|3\rangle$ with the normalization relation $|c_1|^2+|c_2|^2+|c_3|^2=1$ is the evolutive state, one can obtain
\begin{align*}\label{e3a}
&i\dot c_1=\frac{\Omega_p}2\mathcal{C}_2\pm\frac{\Omega_q}2c_3,\nonumber\\
&i\dot{\mathcal{C}}_2=\Delta\mathcal{C}_2+\frac{\Omega_p}2c_1+\frac{\Omega_s}2c_3,\nonumber\\
&i\dot c_3=\frac{\Omega_s}2\mathcal{C}_2\pm\frac{\Omega_q}2c_1,\tag*{(A1)}
\end{align*}
with $\mathcal{C}_2=e^{-i\Delta t}c_2$. A large $\Delta$ makes $|i\dot{\mathcal{C}}_2|\ll|\Delta\mathcal{C}_2|$, thus we can set $\dot{\mathcal{C}}_2=0$ that gives
\begin{align*}\label{ec2}
\mathcal{C}_2=-\frac1\Delta\left(\frac{\Omega_p}2c_1+\frac{\Omega_s}2c_3\right).\tag*{(A2)}
\end{align*}
By substituting Eq.~\ref{ec2} into Eq.~\ref{e3a}, one can get
\begin{align*}\label{c1c3}
i\dot c_1&=-\frac{\Omega_p^2}{4\Delta}{c}_1-\left(\frac{\Omega_p\Omega_s}{4\Delta}\mp\frac{\Omega_q}2\right){c}_3,\nonumber\\
i\dot c_3&=-\frac{\Omega_s^2}{4\Delta}{c}_3-\left(\frac{\Omega_p\Omega_s}{4\Delta}\mp\frac{\Omega_q}2\right){c}_1.\tag*{(A3)}
\end{align*}
Conversely, according to Eq.~\ref{c1c3} an effective Hamiltonian of $\hat H_{L,R}$ in Eq.~(\ref{e2}) can be developed
\begin{align*}\label{A4}
\hat H^e_{L,R}&=\frac{\Omega_p^2}{4\Delta}|1\rangle\langle 1|+\frac{\Omega_s^2}{4\Delta}|3\rangle\langle 3|\nonumber\\
&\quad+\left(\frac{\Omega_p\Omega_s}{4\Delta}\mp\frac{\Omega_q}2\right)(|1\rangle\langle 3|+|3\rangle\langle 1|),\tag*{(A4)}
\end{align*}
for which we have neglected the global phase. The first two terms in Eq.~\ref{A4} are the Stark shifts of $|1\rangle$ and $|3\rangle$, while the last term denotes the effective interaction between $|1\rangle$ and $|3\rangle$ with an effective Rabi frequency $\Omega_{\rm eff}\equiv{\Omega_p\Omega_s}/{2\Delta}$. Further, the Stark-shift terms can be dropped with $\Omega_p=\Omega_s$ because of the completeness $|1\rangle\langle 1|+|3\rangle\langle 3|=\hat{\mathcal{I}}$, $\hat{\mathcal{I}}$ being the unit operator.

\section*{Appendix B: Master equation with unwanted transitions and an imperfect initial state}
The five considered transitions $|1\rangle\leftrightarrow|2\rangle$, $|1\rangle\leftrightarrow|3\rangle$, $|2\rangle\leftrightarrow|3\rangle$, $|1\rangle\leftrightarrow|4\rangle$, and $|2\rangle\leftrightarrow|4\rangle$ are of the transition frequencies $\omega_{12}=7059$~MHz~($b$-type), $\omega_{13}=4720$~MHz~($a$-type), $\omega_{23}=2339$~MHz~($c$-type), $\omega_{14}=2575$~MHz~($c$-type), and $\omega_{24}=4484$~MHz~($a$-type), respectively. The $a$-type, $b$-type, and $c$-type transitions are driven by orthogonal $Q$ pulse at frequency $\omega_q=\omega_{13}$, $P$ pulse at frequency $\omega_p=\omega_{12}-\Delta$, and $S$ pulse at frequency $\omega_s=\omega_{23}-\Delta$, respectively. By choosing appropriate axes such that each transition dipole moment and its driving field has the same orientation, the evolutions of two enantiomers can be governed by the Hamiltonian
\begin{align*}\label{eB1}
\hat H_{L,R}&=\sum_{j=2}^4\omega_{1j}|j\rangle\langle j|+[\Omega_p\cos\omega_pt|1\rangle\langle2|\nonumber\\
&\quad\pm \Omega_q\cos\omega_qt(|1\rangle\langle3|+|2\rangle\langle4|)\nonumber\\
&\quad+ \Omega_s\cos\omega_st(|2\rangle\langle3|+|1\rangle\langle4|)+{\rm H.c.}].\tag*{(B1)}
\end{align*}

An experiment of the present proposal is supposed to carry out with the ultra-cooled cyclohexylmethanol sample, for which the temperature of sample should be low enough so that all molecules can be prepared perfectly in the initial state $|1\rangle$. Buffer gas cooling and supersonic expansions are two valid tools for cooling the molecular samples. In recent experiments, the molecular samples can be cooled to a temperature of around $5$-$10$~K by using a cryogenic buffer gas cell~\cite{Patterson2013,Patterson2013PRL,Eibenberger2017}. The supersonic expansion can cool the molecules to rotational temperatures of about $1$-$2$~K~\cite{Shubert2014,Shubert2015,Perez2017,Perez2018}. Here we assume that the present proposal is carried out at a relatively low temperature, and all molecules are prepared initially in a lowly-mixed state with the density operator $\hat\rho_0=0.998|1\rangle\langle1|+0.001|2\rangle\langle2|+0.001|3\rangle\langle3|$, which means that the initial state contains a spot of $|2\rangle$ and $|3\rangle$ with the same mixed probability $0.001$. Then the time evolution of the density operator $\hat\rho$ can be ruled by the Markovian master equation when the energy relaxations of higher energy states are taken into account
\begin{align*}\label{eB2}
\frac{\partial{\hat \rho}}{\partial t}&=i{\hat \rho} \hat H_{L,R}-i\hat H_{L,R}{\hat \rho}\nonumber\\
&\quad-\sum_{j=2}^4\frac{\gamma_{1,j}}{2}\left({\hat\sigma_{1,j}}^\dag\hat\sigma_{1,j}\hat\rho-2\hat\sigma_{1,j}\hat\rho{\hat\sigma_{1,j}}^\dag+\hat\rho{\hat\sigma_{1,j}}^\dag\hat\sigma_{1,j}\right)
\nonumber\\
&\quad-\sum_{k=3,4}\frac{\gamma_{k,2}}{2}\left({\hat\sigma_{k,2}}^\dag\hat\sigma_{k,2}\hat\rho-2\hat\sigma_{k,2}\hat\rho{\hat\sigma_{k,2}}^\dag+\hat\rho{\hat\sigma_{k,2}^\dag}\hat\sigma_{k,2}\right),\tag*{(B2)}
\end{align*}
with $\gamma_{m,n}$ being the relaxation rate from $|n\rangle$ to $|m\rangle$ and $\hat \sigma_{m,n}\equiv|m\rangle\langle n|$ the relaxation operator.

\section*{Appendix C: Dependence of the enantiomer-selective excitation on field phases}
Dependence of the enantiomer-selective tasks on the phases of fields is a common property for the existing schemes, and also it can be used to switch the excitation of the two enantiomers. Phases of the three microwave fields are chosen as zero for convenience, while it is of course not a sole choice. When taking nonzero phases of the three microwave fields into account, the effective Hamiltonian Eq.~(\ref{e3}) for the two enantiomers becomes
\begin{align*}\label{eC1}
\hat H_{L,R}^{\rm{eff}}=\frac{\Omega_qe^{i\phi_qt}\mp\Omega_{\rm eff}e^{i(\phi_p-\phi_s)t}}2|1\rangle\langle 3|+{\rm H.c.},\tag*{(C1)}
\end{align*}
From Eq.~\ref{eC1} we can find that the three field phases have decisive effect on the two-path interference peculiarities of the two enantiomers. For the interference between the one- and two-photon $|1\rangle\leftrightarrow|3\rangle$ transitions being constructive for one enantiomer but destructive for the other, the condition is
\begin{align*}\label{eC2}
\phi_q-\phi_p+\phi_s=n\pi,\quad n\in{\rm integers},\tag*{(C2)}
\end{align*}
in which when $n$ is odd~(even), the two-path interference is constructive for $L$-handed~($R$-handed) molecules but destructive for $R$-handed~($L$-handed) ones. Therefore, any one of the three fields having a change of a $\pi$ phase will invert the interference properties of the two enantiomers and switch the excitation of the two enantiomers.

\bibliography{apssamp}

\begin{thebibliography}{62}%
\makeatletter
\providecommand \@ifxundefined [1]{%
 \@ifx{#1\undefined}
}%
\providecommand \@ifnum [1]{%
 \ifnum #1\expandafter \@firstoftwo
 \else \expandafter \@secondoftwo
 \fi
}%
\providecommand \@ifx [1]{%
 \ifx #1\expandafter \@firstoftwo
 \else \expandafter \@secondoftwo
 \fi
}%
\providecommand \natexlab [1]{#1}%
\providecommand \enquote  [1]{``#1''}%
\providecommand \bibnamefont  [1]{#1}%
\providecommand \bibfnamefont [1]{#1}%
\providecommand \citenamefont [1]{#1}%
\providecommand \href@noop [0]{\@secondoftwo}%
\providecommand \href [0]{\begingroup \@sanitize@url \@href}%
\providecommand \@href[1]{\@@startlink{#1}\@@href}%
\providecommand \@@href[1]{\endgroup#1\@@endlink}%
\providecommand \@sanitize@url [0]{\catcode `\\12\catcode `\$12\catcode
  `\&12\catcode `\#12\catcode `\^12\catcode `\_12\catcode `\%12\relax}%
\providecommand \@@startlink[1]{}%
\providecommand \@@endlink[0]{}%
\providecommand \url  [0]{\begingroup\@sanitize@url \@url }%
\providecommand \@url [1]{\endgroup\@href {#1}{\urlprefix }}%
\providecommand \urlprefix  [0]{URL }%
\providecommand \Eprint [0]{\href }%
\providecommand \doibase [0]{http://dx.doi.org/}%
\providecommand \selectlanguage [0]{\@gobble}%
\providecommand \bibinfo  [0]{\@secondoftwo}%
\providecommand \bibfield  [0]{\@secondoftwo}%
\providecommand \translation [1]{[#1]}%
\providecommand \BibitemOpen [0]{}%
\providecommand \bibitemStop [0]{}%
\providecommand \bibitemNoStop [0]{.\EOS\space}%
\providecommand \EOS [0]{\spacefactor3000\relax}%
\providecommand \BibitemShut  [1]{\csname bibitem#1\endcsname}%
\let\auto@bib@innerbib\@empty
\bibitem [{\citenamefont {Barrett}\ \emph {et~al.}(2014)\citenamefont
  {Barrett}, \citenamefont {Metrano}, \citenamefont {Rablen},\ and\
  \citenamefont {Miller}}]{Barrett2014}%
  \BibitemOpen
  \bibfield  {author} {\bibinfo {author} {\bibfnamefont {K.~T.}\ \bibnamefont
  {Barrett}}, \bibinfo {author} {\bibfnamefont {A.~J.}\ \bibnamefont
  {Metrano}}, \bibinfo {author} {\bibfnamefont {P.~R.}\ \bibnamefont {Rablen}},
  \ and\ \bibinfo {author} {\bibfnamefont {S.~J.}\ \bibnamefont {Miller}},\
  }\href {\doibase 10.1038/nature13189} {\bibfield  {journal} {\bibinfo
  {journal} {Nature}\ }\textbf {\bibinfo {volume} {509}},\ \bibinfo {pages}
  {71} (\bibinfo {year} {2014})}\BibitemShut {NoStop}%
\bibitem [{\citenamefont {Kaushik}\ \emph {et~al.}(2015)\citenamefont
  {Kaushik}, \citenamefont {Basu}, \citenamefont {Benoit}, \citenamefont
  {Cirtiu}, \citenamefont {Vali},\ and\ \citenamefont {Moores}}]{Kaushik2015}%
  \BibitemOpen
  \bibfield  {author} {\bibinfo {author} {\bibfnamefont {M.}~\bibnamefont
  {Kaushik}}, \bibinfo {author} {\bibfnamefont {K.}~\bibnamefont {Basu}},
  \bibinfo {author} {\bibfnamefont {C.}~\bibnamefont {Benoit}}, \bibinfo
  {author} {\bibfnamefont {C.~M.}\ \bibnamefont {Cirtiu}}, \bibinfo {author}
  {\bibfnamefont {H.}~\bibnamefont {Vali}}, \ and\ \bibinfo {author}
  {\bibfnamefont {A.}~\bibnamefont {Moores}},\ }\href {\doibase
  10.1021/jacs.5b02034} {\bibfield  {journal} {\bibinfo  {journal} {J. Am.
  Chem. Soc.}\ }\textbf {\bibinfo {volume} {137}},\ \bibinfo {pages} {6124}
  (\bibinfo {year} {2015})}\BibitemShut {NoStop}%
\bibitem [{\citenamefont {Jiang}\ \emph {et~al.}(2017)\citenamefont {Jiang},
  \citenamefont {Beiger},\ and\ \citenamefont {Hartwig}}]{Jiang2017}%
  \BibitemOpen
  \bibfield  {author} {\bibinfo {author} {\bibfnamefont {X.}~\bibnamefont
  {Jiang}}, \bibinfo {author} {\bibfnamefont {J.~J.}\ \bibnamefont {Beiger}}, \
  and\ \bibinfo {author} {\bibfnamefont {J.~F.}\ \bibnamefont {Hartwig}},\
  }\href {\doibase 10.1021/jacs.6b11692} {\bibfield  {journal} {\bibinfo
  {journal} {J. Am. Chem. Soc.}\ }\textbf {\bibinfo {volume} {139}},\ \bibinfo
  {pages} {87} (\bibinfo {year} {2017})}\BibitemShut {NoStop}%
\bibitem [{\citenamefont {Gal}(2012)}]{Joseph2012}%
  \BibitemOpen
  \bibfield  {author} {\bibinfo {author} {\bibfnamefont {J.}~\bibnamefont
  {Gal}},\ }\href {\doibase 10.1002/chir.22071} {\bibfield  {journal} {\bibinfo
   {journal} {Chirality}\ }\textbf {\bibinfo {volume} {24}},\ \bibinfo {pages}
  {959} (\bibinfo {year} {2012})}\BibitemShut {NoStop}%
\bibitem [{\citenamefont {Intlekofer}\ \emph {et~al.}(2015)\citenamefont
  {Intlekofer}, \citenamefont {Dematteo}, \citenamefont {Venneti},
  \citenamefont {Finley}, \citenamefont {Lu}, \citenamefont {Judkins},
  \citenamefont {Rustenburg}, \citenamefont {Grinaway}, \citenamefont
  {Chodera}, \citenamefont {Cross},\ and\ \citenamefont
  {Thompson}}]{INTLEKOFER2015304}%
  \BibitemOpen
  \bibfield  {author} {\bibinfo {author} {\bibfnamefont {A.~M.}\ \bibnamefont
  {Intlekofer}}, \bibinfo {author} {\bibfnamefont {R.~G.}\ \bibnamefont
  {Dematteo}}, \bibinfo {author} {\bibfnamefont {S.}~\bibnamefont {Venneti}},
  \bibinfo {author} {\bibfnamefont {L.~W.~S.}\ \bibnamefont {Finley}}, \bibinfo
  {author} {\bibfnamefont {C.}~\bibnamefont {Lu}}, \bibinfo {author}
  {\bibfnamefont {A.~R.}\ \bibnamefont {Judkins}}, \bibinfo {author}
  {\bibfnamefont {A.~S.}\ \bibnamefont {Rustenburg}}, \bibinfo {author}
  {\bibfnamefont {P.~B.}\ \bibnamefont {Grinaway}}, \bibinfo {author}
  {\bibfnamefont {J.~D.}\ \bibnamefont {Chodera}}, \bibinfo {author}
  {\bibfnamefont {J.~R.}\ \bibnamefont {Cross}}, \ and\ \bibinfo {author}
  {\bibfnamefont {C.~B.}\ \bibnamefont {Thompson}},\ }\href {\doibase
  https://doi.org/10.1016/j.cmet.2015.06.023} {\bibfield  {journal} {\bibinfo
  {journal} {Cell Metab.}\ }\textbf {\bibinfo {volume} {22}},\ \bibinfo {pages}
  {304} (\bibinfo {year} {2015})}\BibitemShut {NoStop}%
\bibitem [{\citenamefont {Chen}\ and\ \citenamefont
  {Ma}(2020)}]{10.1371/journal.pcbi.1007592}%
  \BibitemOpen
  \bibfield  {author} {\bibinfo {author} {\bibfnamefont {Y.}~\bibnamefont
  {Chen}}\ and\ \bibinfo {author} {\bibfnamefont {W.}~\bibnamefont {Ma}},\
  }\href {\doibase 10.1371/journal.pcbi.1007592} {\bibfield  {journal}
  {\bibinfo  {journal} {PLOS Comput. Biol.}\ }\textbf {\bibinfo {volume}
  {16}},\ \bibinfo {pages} {1} (\bibinfo {year} {2020})}\BibitemShut {NoStop}%
\bibitem [{\citenamefont {Ma}\ \emph {et~al.}(2016)\citenamefont {Ma},
  \citenamefont {Wang}, \citenamefont {Lu}, \citenamefont {Zhang},
  \citenamefont {Yin}, \citenamefont {Huang}, \citenamefont {Deng},
  \citenamefont {Wang},\ and\ \citenamefont {Yu}}]{MA2016268}%
  \BibitemOpen
  \bibfield  {author} {\bibinfo {author} {\bibfnamefont {R.}~\bibnamefont
  {Ma}}, \bibinfo {author} {\bibfnamefont {B.}~\bibnamefont {Wang}}, \bibinfo
  {author} {\bibfnamefont {S.}~\bibnamefont {Lu}}, \bibinfo {author}
  {\bibfnamefont {Y.}~\bibnamefont {Zhang}}, \bibinfo {author} {\bibfnamefont
  {L.}~\bibnamefont {Yin}}, \bibinfo {author} {\bibfnamefont {J.}~\bibnamefont
  {Huang}}, \bibinfo {author} {\bibfnamefont {S.}~\bibnamefont {Deng}},
  \bibinfo {author} {\bibfnamefont {Y.}~\bibnamefont {Wang}}, \ and\ \bibinfo
  {author} {\bibfnamefont {G.}~\bibnamefont {Yu}},\ }\href {\doibase
  https://doi.org/10.1016/j.scitotenv.2016.03.053} {\bibfield  {journal}
  {\bibinfo  {journal} {Sci. Total Environ.}\ }\textbf {\bibinfo {volume}
  {557-558}},\ \bibinfo {pages} {268} (\bibinfo {year} {2016})}\BibitemShut
  {NoStop}%
\bibitem [{\citenamefont {Amorim}\ \emph {et~al.}(2016)\citenamefont {Amorim},
  \citenamefont {Moreira}, \citenamefont {Ribeiro}, \citenamefont {Santos},
  \citenamefont {Delerue-Matos}, \citenamefont {Tiritan},\ and\ \citenamefont
  {Castro}}]{AMORIM2016277}%
  \BibitemOpen
  \bibfield  {author} {\bibinfo {author} {\bibfnamefont {C.~L.}\ \bibnamefont
  {Amorim}}, \bibinfo {author} {\bibfnamefont {I.~S.}\ \bibnamefont {Moreira}},
  \bibinfo {author} {\bibfnamefont {A.~R.}\ \bibnamefont {Ribeiro}}, \bibinfo
  {author} {\bibfnamefont {L.~H.}\ \bibnamefont {Santos}}, \bibinfo {author}
  {\bibfnamefont {C.}~\bibnamefont {Delerue-Matos}}, \bibinfo {author}
  {\bibfnamefont {M.~E.}\ \bibnamefont {Tiritan}}, \ and\ \bibinfo {author}
  {\bibfnamefont {P.~M.}\ \bibnamefont {Castro}},\ }\href {\doibase
  https://doi.org/10.1016/j.ibiod.2016.09.009} {\bibfield  {journal} {\bibinfo
  {journal} {Int. Biodeter. Biodegr.}\ }\textbf {\bibinfo {volume} {115}},\
  \bibinfo {pages} {277} (\bibinfo {year} {2016})}\BibitemShut {NoStop}%
\bibitem [{\citenamefont {Sanganyado}\ \emph {et~al.}(2017)\citenamefont
  {Sanganyado}, \citenamefont {Lu}, \citenamefont {Fu}, \citenamefont
  {Schlenk},\ and\ \citenamefont {Gan}}]{SANGANYADO2017527}%
  \BibitemOpen
  \bibfield  {author} {\bibinfo {author} {\bibfnamefont {E.}~\bibnamefont
  {Sanganyado}}, \bibinfo {author} {\bibfnamefont {Z.}~\bibnamefont {Lu}},
  \bibinfo {author} {\bibfnamefont {Q.}~\bibnamefont {Fu}}, \bibinfo {author}
  {\bibfnamefont {D.}~\bibnamefont {Schlenk}}, \ and\ \bibinfo {author}
  {\bibfnamefont {J.}~\bibnamefont {Gan}},\ }\href {\doibase
  https://doi.org/10.1016/j.watres.2017.08.003} {\bibfield  {journal} {\bibinfo
   {journal} {Water Res.}\ }\textbf {\bibinfo {volume} {124}},\ \bibinfo
  {pages} {527} (\bibinfo {year} {2017})}\BibitemShut {NoStop}%
\bibitem [{\citenamefont {Ribeiro}\ \emph {et~al.}(2020)\citenamefont
  {Ribeiro}, \citenamefont {Castro},\ and\ \citenamefont
  {Tiritan}}]{Ribeiro2020}%
  \BibitemOpen
  \bibfield  {author} {\bibinfo {author} {\bibfnamefont {A.~R.}\ \bibnamefont
  {Ribeiro}}, \bibinfo {author} {\bibfnamefont {P.~M.~L.}\ \bibnamefont
  {Castro}}, \ and\ \bibinfo {author} {\bibfnamefont {M.~E.}\ \bibnamefont
  {Tiritan}},\ }\href {\doibase 10.1371/journal.pcbi.1007592} {\bibfield
  {journal} {\bibinfo  {journal} {Environ. Chem. Lett.}\ }\textbf {\bibinfo
  {volume} {16}},\ \bibinfo {pages} {1} (\bibinfo {year} {2020})}\BibitemShut
  {NoStop}%
\bibitem [{\citenamefont {Bodenh{\"o}fer}\ \emph {et~al.}(1997)\citenamefont
  {Bodenh{\"o}fer}, \citenamefont {Hierlemann}, \citenamefont {Seemann},
  \citenamefont {Gauglitz}, \citenamefont {Koppenhoefer},\ and\ \citenamefont
  {Gpel}}]{Boden}%
  \BibitemOpen
  \bibfield  {author} {\bibinfo {author} {\bibfnamefont {K.}~\bibnamefont
  {Bodenh{\"o}fer}}, \bibinfo {author} {\bibfnamefont {A.}~\bibnamefont
  {Hierlemann}}, \bibinfo {author} {\bibfnamefont {J.}~\bibnamefont {Seemann}},
  \bibinfo {author} {\bibfnamefont {G.}~\bibnamefont {Gauglitz}}, \bibinfo
  {author} {\bibfnamefont {B.}~\bibnamefont {Koppenhoefer}}, \ and\ \bibinfo
  {author} {\bibfnamefont {W.}~\bibnamefont {Gpel}},\ }\href {\doibase
  10.1038/42426} {\bibfield  {journal} {\bibinfo  {journal} {Nature}\ }\textbf
  {\bibinfo {volume} {387}},\ \bibinfo {pages} {577} (\bibinfo {year}
  {1997})}\BibitemShut {NoStop}%
\bibitem [{\citenamefont {McKendry}\ \emph {et~al.}(1998)\citenamefont
  {McKendry}, \citenamefont {Theoclitou}, \citenamefont {Rayment},\ and\
  \citenamefont {Abell}}]{McKendry}%
  \BibitemOpen
  \bibfield  {author} {\bibinfo {author} {\bibfnamefont {R.}~\bibnamefont
  {McKendry}}, \bibinfo {author} {\bibfnamefont {M.-E.}\ \bibnamefont
  {Theoclitou}}, \bibinfo {author} {\bibfnamefont {T.}~\bibnamefont {Rayment}},
  \ and\ \bibinfo {author} {\bibfnamefont {C.}~\bibnamefont {Abell}},\ }\href
  {\doibase 110.1038/35339} {\bibfield  {journal} {\bibinfo  {journal}
  {Nature}\ }\textbf {\bibinfo {volume} {391}},\ \bibinfo {pages} {566}
  (\bibinfo {year} {1998})}\BibitemShut {NoStop}%
\bibitem [{\citenamefont {Rikken}\ and\ \citenamefont
  {Raupach}(2000)}]{Rikken}%
  \BibitemOpen
  \bibfield  {author} {\bibinfo {author} {\bibfnamefont {G.}~\bibnamefont
  {Rikken}}\ and\ \bibinfo {author} {\bibfnamefont {E.}~\bibnamefont
  {Raupach}},\ }\href {\doibase 10.1038/35016043} {\bibfield  {journal}
  {\bibinfo  {journal} {Nature}\ }\textbf {\bibinfo {volume} {405}},\ \bibinfo
  {pages} {932} (\bibinfo {year} {2000})}\BibitemShut {NoStop}%
\bibitem [{\citenamefont {Zepik}\ \emph {et~al.}(2002)\citenamefont {Zepik},
  \citenamefont {Shavit}, \citenamefont {Tang}, \citenamefont {Jensen},
  \citenamefont {Kjaer}, \citenamefont {Bolbach}, \citenamefont {Leiserowitz},
  \citenamefont {Weissbuch},\ and\ \citenamefont {Lahav}}]{Zepik}%
  \BibitemOpen
  \bibfield  {author} {\bibinfo {author} {\bibfnamefont {H.}~\bibnamefont
  {Zepik}}, \bibinfo {author} {\bibfnamefont {E.}~\bibnamefont {Shavit}},
  \bibinfo {author} {\bibfnamefont {M.}~\bibnamefont {Tang}}, \bibinfo {author}
  {\bibfnamefont {T.~R.}\ \bibnamefont {Jensen}}, \bibinfo {author}
  {\bibfnamefont {K.}~\bibnamefont {Kjaer}}, \bibinfo {author} {\bibfnamefont
  {G.}~\bibnamefont {Bolbach}}, \bibinfo {author} {\bibfnamefont
  {L.}~\bibnamefont {Leiserowitz}}, \bibinfo {author} {\bibfnamefont
  {I.}~\bibnamefont {Weissbuch}}, \ and\ \bibinfo {author} {\bibfnamefont
  {M.}~\bibnamefont {Lahav}},\ }\href {\doibase 10.1126/science.1065625}
  {\bibfield  {journal} {\bibinfo  {journal} {Science}\ }\textbf {\bibinfo
  {volume} {295}},\ \bibinfo {pages} {1266} (\bibinfo {year}
  {2002})}\BibitemShut {NoStop}%
\bibitem [{\citenamefont {Bielski}\ and\ \citenamefont
  {Tencer}(2005)}]{Bielski1}%
  \BibitemOpen
  \bibfield  {author} {\bibinfo {author} {\bibfnamefont {R.}~\bibnamefont
  {Bielski}}\ and\ \bibinfo {author} {\bibfnamefont {M.}~\bibnamefont
  {Tencer}},\ }\href {\doibase 10.1002/jssc.200500173} {\bibfield  {journal}
  {\bibinfo  {journal} {J. Sep. Sci.}\ }\textbf {\bibinfo {volume} {28}},\
  \bibinfo {pages} {2325} (\bibinfo {year} {2005})}\BibitemShut {NoStop}%
\bibitem [{\citenamefont {Bielski}\ and\ \citenamefont
  {Tencer}(2007)}]{Bielski2}%
  \BibitemOpen
  \bibfield  {author} {\bibinfo {author} {\bibfnamefont {R.}~\bibnamefont
  {Bielski}}\ and\ \bibinfo {author} {\bibfnamefont {M.}~\bibnamefont
  {Tencer}},\ }\href {\doibase 10.1007/s11084-006-9022-9} {\bibfield  {journal}
  {\bibinfo  {journal} {Origins Life Evol. Biosphere}\ }\textbf {\bibinfo
  {volume} {37}},\ \bibinfo {pages} {167} (\bibinfo {year} {2007})}\BibitemShut
  {NoStop}%
\bibitem [{\citenamefont {Cameron}\ \emph {et~al.}(2014)\citenamefont
  {Cameron}, \citenamefont {Barnett},\ and\ \citenamefont
  {Yao}}]{Cameron_2014}%
  \BibitemOpen
  \bibfield  {author} {\bibinfo {author} {\bibfnamefont {R.~P.}\ \bibnamefont
  {Cameron}}, \bibinfo {author} {\bibfnamefont {S.~M.}\ \bibnamefont
  {Barnett}}, \ and\ \bibinfo {author} {\bibfnamefont {A.~M.}\ \bibnamefont
  {Yao}},\ }\href {\doibase 10.1088/1367-2630/16/1/013020} {\bibfield
  {journal} {\bibinfo  {journal} {New J. Phys.}\ }\textbf {\bibinfo {volume}
  {16}},\ \bibinfo {pages} {013020} (\bibinfo {year} {2014})}\BibitemShut
  {NoStop}%
\bibitem [{\citenamefont {Bradshaw}\ \emph {et~al.}(2015)\citenamefont
  {Bradshaw}, \citenamefont {Forbes}, \citenamefont {Leeder},\ and\
  \citenamefont {Andrews}}]{Bradshaw2015}%
  \BibitemOpen
  \bibfield  {author} {\bibinfo {author} {\bibfnamefont {D.~S.}\ \bibnamefont
  {Bradshaw}}, \bibinfo {author} {\bibfnamefont {K.~A.}\ \bibnamefont
  {Forbes}}, \bibinfo {author} {\bibfnamefont {J.~M.}\ \bibnamefont {Leeder}},
  \ and\ \bibinfo {author} {\bibfnamefont {D.~L.}\ \bibnamefont {Andrews}},\
  }\href {\doibase 10.3390/photonics2020483} {\bibfield  {journal} {\bibinfo
  {journal} {Photonics}\ }\textbf {\bibinfo {volume} {2}},\ \bibinfo {pages}
  {483} (\bibinfo {year} {2015})}\BibitemShut {NoStop}%
\bibitem [{\citenamefont {Bradshaw}\ and\ \citenamefont
  {Andrews}(2015)}]{Bradshaw:15}%
  \BibitemOpen
  \bibfield  {author} {\bibinfo {author} {\bibfnamefont {D.~S.}\ \bibnamefont
  {Bradshaw}}\ and\ \bibinfo {author} {\bibfnamefont {D.~L.}\ \bibnamefont
  {Andrews}},\ }\href {\doibase 10.1364/OL.40.000677} {\bibfield  {journal}
  {\bibinfo  {journal} {Opt. Lett.}\ }\textbf {\bibinfo {volume} {40}},\
  \bibinfo {pages} {677} (\bibinfo {year} {2015})}\BibitemShut {NoStop}%
\bibitem [{\citenamefont {Cameron}\ \emph {et~al.}(2016)\citenamefont
  {Cameron}, \citenamefont {G\"otte},\ and\ \citenamefont
  {Barnett}}]{PhysRevA.94.032505}%
  \BibitemOpen
  \bibfield  {author} {\bibinfo {author} {\bibfnamefont {R.~P.}\ \bibnamefont
  {Cameron}}, \bibinfo {author} {\bibfnamefont {J.~B.}\ \bibnamefont
  {G\"otte}}, \ and\ \bibinfo {author} {\bibfnamefont {S.~M.}\ \bibnamefont
  {Barnett}},\ }\href {\doibase 10.1103/PhysRevA.94.032505} {\bibfield
  {journal} {\bibinfo  {journal} {Phys. Rev. A}\ }\textbf {\bibinfo {volume}
  {94}},\ \bibinfo {pages} {032505} (\bibinfo {year} {2016})}\BibitemShut
  {NoStop}%
\bibitem [{\citenamefont {Milner}\ \emph {et~al.}(2019)\citenamefont {Milner},
  \citenamefont {Fordyce}, \citenamefont {MacPhail-Bartley}, \citenamefont
  {Wasserman}, \citenamefont {Milner}, \citenamefont {Tutunnikov},\ and\
  \citenamefont {Averbukh}}]{PhysRevLett.122.223201}%
  \BibitemOpen
  \bibfield  {author} {\bibinfo {author} {\bibfnamefont {A.~A.}\ \bibnamefont
  {Milner}}, \bibinfo {author} {\bibfnamefont {J.~A.~M.}\ \bibnamefont
  {Fordyce}}, \bibinfo {author} {\bibfnamefont {I.}~\bibnamefont
  {MacPhail-Bartley}}, \bibinfo {author} {\bibfnamefont {W.}~\bibnamefont
  {Wasserman}}, \bibinfo {author} {\bibfnamefont {V.}~\bibnamefont {Milner}},
  \bibinfo {author} {\bibfnamefont {I.}~\bibnamefont {Tutunnikov}}, \ and\
  \bibinfo {author} {\bibfnamefont {I.~S.}\ \bibnamefont {Averbukh}},\ }\href
  {\doibase 10.1103/PhysRevLett.122.223201} {\bibfield  {journal} {\bibinfo
  {journal} {Phys. Rev. Lett.}\ }\textbf {\bibinfo {volume} {122}},\ \bibinfo
  {pages} {223201} (\bibinfo {year} {2019})}\BibitemShut {NoStop}%
\bibitem [{\citenamefont {Yachmenev}\ \emph {et~al.}(2019)\citenamefont
  {Yachmenev}, \citenamefont {Onvlee}, \citenamefont {Zak}, \citenamefont
  {Owens},\ and\ \citenamefont {K\"upper}}]{PhysRevLett.123.243202}%
  \BibitemOpen
  \bibfield  {author} {\bibinfo {author} {\bibfnamefont {A.}~\bibnamefont
  {Yachmenev}}, \bibinfo {author} {\bibfnamefont {J.}~\bibnamefont {Onvlee}},
  \bibinfo {author} {\bibfnamefont {E.}~\bibnamefont {Zak}}, \bibinfo {author}
  {\bibfnamefont {A.}~\bibnamefont {Owens}}, \ and\ \bibinfo {author}
  {\bibfnamefont {J.}~\bibnamefont {K\"upper}},\ }\href {\doibase
  10.1103/PhysRevLett.123.243202} {\bibfield  {journal} {\bibinfo  {journal}
  {Phys. Rev. Lett.}\ }\textbf {\bibinfo {volume} {123}},\ \bibinfo {pages}
  {243202} (\bibinfo {year} {2019})}\BibitemShut {NoStop}%
\bibitem [{\citenamefont {Berova}\ \emph {et~al.}(2007)\citenamefont {Berova},
  \citenamefont {Bari},\ and\ \citenamefont {Pescitelli}}]{B515476F}%
  \BibitemOpen
  \bibfield  {author} {\bibinfo {author} {\bibfnamefont {N.}~\bibnamefont
  {Berova}}, \bibinfo {author} {\bibfnamefont {L.~D.}\ \bibnamefont {Bari}}, \
  and\ \bibinfo {author} {\bibfnamefont {G.}~\bibnamefont {Pescitelli}},\
  }\href {\doibase 10.1039/B515476F} {\bibfield  {journal} {\bibinfo  {journal}
  {Chem. Soc. Rev.}\ }\textbf {\bibinfo {volume} {36}},\ \bibinfo {pages} {914}
  (\bibinfo {year} {2007})}\BibitemShut {NoStop}%
\bibitem [{\citenamefont {Nafie}(2011)}]{Nafie2011}%
  \BibitemOpen
  \bibfield  {author} {\bibinfo {author} {\bibfnamefont {L.~A.}\ \bibnamefont
  {Nafie}},\ }\href@noop {} {\emph {\bibinfo {title} {Vibrational Optical
  Activity}}}\ (\bibinfo  {publisher} {Wiley, New York},\ \bibinfo {year}
  {2011})\BibitemShut {NoStop}%
\bibitem [{\citenamefont {Barron}\ \emph {et~al.}(2007)\citenamefont {Barron},
  \citenamefont {Zhu}, \citenamefont {Hecht}, \citenamefont {Tranter},\ and\
  \citenamefont {Isaacs}}]{BARRON20077}%
  \BibitemOpen
  \bibfield  {author} {\bibinfo {author} {\bibfnamefont {L.~D.}\ \bibnamefont
  {Barron}}, \bibinfo {author} {\bibfnamefont {F.}~\bibnamefont {Zhu}},
  \bibinfo {author} {\bibfnamefont {L.}~\bibnamefont {Hecht}}, \bibinfo
  {author} {\bibfnamefont {G.~E.}\ \bibnamefont {Tranter}}, \ and\ \bibinfo
  {author} {\bibfnamefont {N.~W.}\ \bibnamefont {Isaacs}},\ }\href {\doibase
  https://doi.org/10.1016/j.molstruc.2006.10.033} {\bibfield  {journal}
  {\bibinfo  {journal} {J. Mol. Struct.}\ }\textbf {\bibinfo {volume}
  {834-836}},\ \bibinfo {pages} {7 } (\bibinfo {year} {2007})}\BibitemShut
  {NoStop}%
\bibitem [{\citenamefont {Barron}(2004)}]{Barron2004}%
  \BibitemOpen
  \bibfield  {author} {\bibinfo {author} {\bibfnamefont {L.~D.}\ \bibnamefont
  {Barron}},\ }\href@noop {} {\emph {\bibinfo {title} {Molecular light
  scattering and optical activity}}}\ (\bibinfo  {publisher} {Cambridge
  University Press},\ \bibinfo {year} {2004})\BibitemShut {NoStop}%
\bibitem [{\citenamefont {Nafie}(2013)}]{Nafie2013}%
  \BibitemOpen
  \bibfield  {author} {\bibinfo {author} {\bibfnamefont {L.~A.}\ \bibnamefont
  {Nafie}},\ }\href {\doibase 10.1038/497446b} {\bibfield  {journal} {\bibinfo
  {journal} {Nature}\ }\textbf {\bibinfo {volume} {497}},\ \bibinfo {pages}
  {446} (\bibinfo {year} {2013})}\BibitemShut {NoStop}%
\bibitem [{\citenamefont {Patterson}\ and\ \citenamefont
  {Schnell}(2014)}]{Patterson2014}%
  \BibitemOpen
  \bibfield  {author} {\bibinfo {author} {\bibfnamefont {D.}~\bibnamefont
  {Patterson}}\ and\ \bibinfo {author} {\bibfnamefont {M.}~\bibnamefont
  {Schnell}},\ }\href {\doibase 10.1039/C4CP00417E} {\bibfield  {journal}
  {\bibinfo  {journal} {Phys. Chem. Chem. Phys.}\ }\textbf {\bibinfo {volume}
  {16}},\ \bibinfo {pages} {11114} (\bibinfo {year} {2014})}\BibitemShut
  {NoStop}%
\bibitem [{\citenamefont {Rouxel}\ \emph {et~al.}(2017)\citenamefont {Rouxel},
  \citenamefont {Kowalewski},\ and\ \citenamefont
  {Mukamel}}]{doi:10.1063/1.4974260}%
  \BibitemOpen
  \bibfield  {author} {\bibinfo {author} {\bibfnamefont {J.~R.}\ \bibnamefont
  {Rouxel}}, \bibinfo {author} {\bibfnamefont {M.}~\bibnamefont {Kowalewski}},
  \ and\ \bibinfo {author} {\bibfnamefont {S.}~\bibnamefont {Mukamel}},\ }\href
  {\doibase 10.1063/1.4974260} {\bibfield  {journal} {\bibinfo  {journal}
  {Struct. Dyn.}\ }\textbf {\bibinfo {volume} {4}},\ \bibinfo {pages} {044006}
  (\bibinfo {year} {2017})}\BibitemShut {NoStop}%
\bibitem [{\citenamefont {Rouxel}\ \emph {et~al.}(2018)\citenamefont {Rouxel},
  \citenamefont {Kowalewski},\ and\ \citenamefont {Mukamel}}]{Rouxel2018}%
  \BibitemOpen
  \bibfield  {author} {\bibinfo {author} {\bibfnamefont {J.~R.}\ \bibnamefont
  {Rouxel}}, \bibinfo {author} {\bibfnamefont {M.}~\bibnamefont {Kowalewski}},
  \ and\ \bibinfo {author} {\bibfnamefont {S.}~\bibnamefont {Mukamel}},\ }\href
  {\doibase 10.1021/acs.jpclett.8b01095} {\bibfield  {journal} {\bibinfo
  {journal} {J. Phys. Chem. Lett.}\ }\textbf {\bibinfo {volume} {9}},\ \bibinfo
  {pages} {3392} (\bibinfo {year} {2018})}\BibitemShut {NoStop}%
\bibitem [{\citenamefont {Hiramatsu}\ \emph {et~al.}(2013)\citenamefont
  {Hiramatsu}, \citenamefont {Kano},\ and\ \citenamefont
  {Nagata}}]{Hiramatsu:13}%
  \BibitemOpen
  \bibfield  {author} {\bibinfo {author} {\bibfnamefont {K.}~\bibnamefont
  {Hiramatsu}}, \bibinfo {author} {\bibfnamefont {H.}~\bibnamefont {Kano}}, \
  and\ \bibinfo {author} {\bibfnamefont {T.}~\bibnamefont {Nagata}},\ }\href
  {\doibase 10.1364/OE.21.013515} {\bibfield  {journal} {\bibinfo  {journal}
  {Opt. Express}\ }\textbf {\bibinfo {volume} {21}},\ \bibinfo {pages} {13515}
  (\bibinfo {year} {2013})}\BibitemShut {NoStop}%
\bibitem [{\citenamefont {Begzjav}\ \emph {et~al.}(2019)\citenamefont
  {Begzjav}, \citenamefont {Zhang}, \citenamefont {Scully},\ and\ \citenamefont
  {Agarwal}}]{Begzjav:19}%
  \BibitemOpen
  \bibfield  {author} {\bibinfo {author} {\bibfnamefont {T.~K.}\ \bibnamefont
  {Begzjav}}, \bibinfo {author} {\bibfnamefont {Z.}~\bibnamefont {Zhang}},
  \bibinfo {author} {\bibfnamefont {M.~O.}\ \bibnamefont {Scully}}, \ and\
  \bibinfo {author} {\bibfnamefont {G.~S.}\ \bibnamefont {Agarwal}},\ }\href
  {\doibase 10.1364/OE.27.013965} {\bibfield  {journal} {\bibinfo  {journal}
  {Opt. Express}\ }\textbf {\bibinfo {volume} {27}},\ \bibinfo {pages} {13965}
  (\bibinfo {year} {2019})}\BibitemShut {NoStop}%
\bibitem [{\citenamefont {Kr\'al}\ and\ \citenamefont
  {Shapiro}(2001)}]{Kral2001}%
  \BibitemOpen
  \bibfield  {author} {\bibinfo {author} {\bibfnamefont {P.}~\bibnamefont
  {Kr\'al}}\ and\ \bibinfo {author} {\bibfnamefont {M.}~\bibnamefont
  {Shapiro}},\ }\href {\doibase 10.1103/PhysRevLett.87.183002} {\bibfield
  {journal} {\bibinfo  {journal} {Phys. Rev. Lett.}\ }\textbf {\bibinfo
  {volume} {87}},\ \bibinfo {pages} {183002} (\bibinfo {year}
  {2001})}\BibitemShut {NoStop}%
\bibitem [{\citenamefont {Kr\'al}\ \emph {et~al.}(2003)\citenamefont {Kr\'al},
  \citenamefont {Thanopulos}, \citenamefont {Shapiro},\ and\ \citenamefont
  {Cohen}}]{Kral2003}%
  \BibitemOpen
  \bibfield  {author} {\bibinfo {author} {\bibfnamefont {P.}~\bibnamefont
  {Kr\'al}}, \bibinfo {author} {\bibfnamefont {I.}~\bibnamefont {Thanopulos}},
  \bibinfo {author} {\bibfnamefont {M.}~\bibnamefont {Shapiro}}, \ and\
  \bibinfo {author} {\bibfnamefont {D.}~\bibnamefont {Cohen}},\ }\href
  {\doibase 10.1103/PhysRevLett.90.033001} {\bibfield  {journal} {\bibinfo
  {journal} {Phys. Rev. Lett.}\ }\textbf {\bibinfo {volume} {90}},\ \bibinfo
  {pages} {033001} (\bibinfo {year} {2003})}\BibitemShut {NoStop}%
\bibitem [{\citenamefont {Kr\'al}\ \emph {et~al.}(2007)\citenamefont {Kr\'al},
  \citenamefont {Thanopulos},\ and\ \citenamefont {Shapiro}}]{Kral2007}%
  \BibitemOpen
  \bibfield  {author} {\bibinfo {author} {\bibfnamefont {P.}~\bibnamefont
  {Kr\'al}}, \bibinfo {author} {\bibfnamefont {I.}~\bibnamefont {Thanopulos}},
  \ and\ \bibinfo {author} {\bibfnamefont {M.}~\bibnamefont {Shapiro}},\ }\href
  {\doibase 10.1103/RevModPhys.79.53} {\bibfield  {journal} {\bibinfo
  {journal} {Rev. Mod. Phys.}\ }\textbf {\bibinfo {volume} {79}},\ \bibinfo
  {pages} {53} (\bibinfo {year} {2007})}\BibitemShut {NoStop}%
\bibitem [{\citenamefont {Bergmann}\ \emph {et~al.}(1998)\citenamefont
  {Bergmann}, \citenamefont {Theuer},\ and\ \citenamefont
  {Shore}}]{Bergmann1998}%
  \BibitemOpen
  \bibfield  {author} {\bibinfo {author} {\bibfnamefont {K.}~\bibnamefont
  {Bergmann}}, \bibinfo {author} {\bibfnamefont {H.}~\bibnamefont {Theuer}}, \
  and\ \bibinfo {author} {\bibfnamefont {B.~W.}\ \bibnamefont {Shore}},\ }\href
  {\doibase 10.1103/RevModPhys.70.1003} {\bibfield  {journal} {\bibinfo
  {journal} {Rev. Mod. Phys.}\ }\textbf {\bibinfo {volume} {70}},\ \bibinfo
  {pages} {1003} (\bibinfo {year} {1998})}\BibitemShut {NoStop}%
\bibitem [{\citenamefont {Vitanov}\ \emph {et~al.}(2017)\citenamefont
  {Vitanov}, \citenamefont {Rangelov}, \citenamefont {Shore},\ and\
  \citenamefont {Bergmann}}]{Vitanov2017}%
  \BibitemOpen
  \bibfield  {author} {\bibinfo {author} {\bibfnamefont {N.~V.}\ \bibnamefont
  {Vitanov}}, \bibinfo {author} {\bibfnamefont {A.~A.}\ \bibnamefont
  {Rangelov}}, \bibinfo {author} {\bibfnamefont {B.~W.}\ \bibnamefont {Shore}},
  \ and\ \bibinfo {author} {\bibfnamefont {K.}~\bibnamefont {Bergmann}},\
  }\href {\doibase 10.1103/RevModPhys.89.015006} {\bibfield  {journal}
  {\bibinfo  {journal} {Rev. Mod. Phys.}\ }\textbf {\bibinfo {volume} {89}},\
  \bibinfo {pages} {015006} (\bibinfo {year} {2017})}\BibitemShut {NoStop}%
\bibitem [{\citenamefont {Li}\ and\ \citenamefont {Bruder}(2008)}]{Li2008}%
  \BibitemOpen
  \bibfield  {author} {\bibinfo {author} {\bibfnamefont {Y.}~\bibnamefont
  {Li}}\ and\ \bibinfo {author} {\bibfnamefont {C.}~\bibnamefont {Bruder}},\
  }\href {\doibase 10.1103/PhysRevA.77.015403} {\bibfield  {journal} {\bibinfo
  {journal} {Phys. Rev. A}\ }\textbf {\bibinfo {volume} {77}},\ \bibinfo
  {pages} {015403} (\bibinfo {year} {2008})}\BibitemShut {NoStop}%
\bibitem [{\citenamefont {Jia}\ and\ \citenamefont {Wei}(2010)}]{Jia_2010}%
  \BibitemOpen
  \bibfield  {author} {\bibinfo {author} {\bibfnamefont {W.~Z.}\ \bibnamefont
  {Jia}}\ and\ \bibinfo {author} {\bibfnamefont {L.~F.}\ \bibnamefont {Wei}},\
  }\href {\doibase 10.1088/0953-4075/43/18/185402} {\bibfield  {journal}
  {\bibinfo  {journal} {J. Phys. B: At. Mol. Opt. Phys.}\ }\textbf {\bibinfo
  {volume} {43}},\ \bibinfo {pages} {185402} (\bibinfo {year}
  {2010})}\BibitemShut {NoStop}%
\bibitem [{\citenamefont {Ye}\ \emph {et~al.}(2019{\natexlab{a}})\citenamefont
  {Ye}, \citenamefont {Zhang}, \citenamefont {Chen},\ and\ \citenamefont
  {Li}}]{PhysRevA.100.033411}%
  \BibitemOpen
  \bibfield  {author} {\bibinfo {author} {\bibfnamefont {C.}~\bibnamefont
  {Ye}}, \bibinfo {author} {\bibfnamefont {Q.}~\bibnamefont {Zhang}}, \bibinfo
  {author} {\bibfnamefont {Y.-Y.}\ \bibnamefont {Chen}}, \ and\ \bibinfo
  {author} {\bibfnamefont {Y.}~\bibnamefont {Li}},\ }\href {\doibase
  10.1103/PhysRevA.100.033411} {\bibfield  {journal} {\bibinfo  {journal}
  {Phys. Rev. A}\ }\textbf {\bibinfo {volume} {100}},\ \bibinfo {pages}
  {033411} (\bibinfo {year} {2019}{\natexlab{a}})}\BibitemShut {NoStop}%
\bibitem [{\citenamefont {Patterson}\ \emph {et~al.}(2013)\citenamefont
  {Patterson}, \citenamefont {Schnell},\ and\ \citenamefont
  {Doyle}}]{Patterson2013}%
  \BibitemOpen
  \bibfield  {author} {\bibinfo {author} {\bibfnamefont {D.}~\bibnamefont
  {Patterson}}, \bibinfo {author} {\bibfnamefont {M.}~\bibnamefont {Schnell}},
  \ and\ \bibinfo {author} {\bibfnamefont {J.~M.}\ \bibnamefont {Doyle}},\
  }\href {\doibase 10.1038/nature12150} {\bibfield  {journal} {\bibinfo
  {journal} {Nature}\ }\textbf {\bibinfo {volume} {497}},\ \bibinfo {pages}
  {475} (\bibinfo {year} {2013})}\BibitemShut {NoStop}%
\bibitem [{\citenamefont {Patterson}\ and\ \citenamefont
  {Doyle}(2013)}]{Patterson2013PRL}%
  \BibitemOpen
  \bibfield  {author} {\bibinfo {author} {\bibfnamefont {D.}~\bibnamefont
  {Patterson}}\ and\ \bibinfo {author} {\bibfnamefont {J.~M.}\ \bibnamefont
  {Doyle}},\ }\href {\doibase 10.1103/PhysRevLett.111.023008} {\bibfield
  {journal} {\bibinfo  {journal} {Phys. Rev. Lett.}\ }\textbf {\bibinfo
  {volume} {111}},\ \bibinfo {pages} {023008} (\bibinfo {year}
  {2013})}\BibitemShut {NoStop}%
\bibitem [{\citenamefont {Shubert}\ \emph {et~al.}(2014)\citenamefont
  {Shubert}, \citenamefont {Schmitz}, \citenamefont {Patterson}, \citenamefont
  {Doyle},\ and\ \citenamefont {Schnell}}]{Shubert2014}%
  \BibitemOpen
  \bibfield  {author} {\bibinfo {author} {\bibfnamefont {V.~A.}\ \bibnamefont
  {Shubert}}, \bibinfo {author} {\bibfnamefont {D.}~\bibnamefont {Schmitz}},
  \bibinfo {author} {\bibfnamefont {D.}~\bibnamefont {Patterson}}, \bibinfo
  {author} {\bibfnamefont {J.~M.}\ \bibnamefont {Doyle}}, \ and\ \bibinfo
  {author} {\bibfnamefont {M.}~\bibnamefont {Schnell}},\ }\href {\doibase
  10.1002/anie.201306271} {\bibfield  {journal} {\bibinfo  {journal} {Angew.
  Chem. Int. Ed.}\ }\textbf {\bibinfo {volume} {53}},\ \bibinfo {pages} {1152}
  (\bibinfo {year} {2014})}\BibitemShut {NoStop}%
\bibitem [{\citenamefont {Lobsiger}\ \emph {et~al.}(2015)\citenamefont
  {Lobsiger}, \citenamefont {P\'{e}rez}, \citenamefont {Evangelisti},
  \citenamefont {Lehmann},\ and\ \citenamefont {Pate}}]{Lobsiger2015}%
  \BibitemOpen
  \bibfield  {author} {\bibinfo {author} {\bibfnamefont {S.}~\bibnamefont
  {Lobsiger}}, \bibinfo {author} {\bibfnamefont {C.}~\bibnamefont {P\'{e}rez}},
  \bibinfo {author} {\bibfnamefont {L.}~\bibnamefont {Evangelisti}}, \bibinfo
  {author} {\bibfnamefont {K.~K.}\ \bibnamefont {Lehmann}}, \ and\ \bibinfo
  {author} {\bibfnamefont {B.~H.}\ \bibnamefont {Pate}},\ }\href {\doibase
  10.1021/jz502312t} {\bibfield  {journal} {\bibinfo  {journal} {J. Phys. Chem.
  Lett.}\ }\textbf {\bibinfo {volume} {6}},\ \bibinfo {pages} {196} (\bibinfo
  {year} {2015})}\BibitemShut {NoStop}%
\bibitem [{\citenamefont {Shubert}\ \emph {et~al.}(2015)\citenamefont
  {Shubert}, \citenamefont {Schmitz}, \citenamefont {Medcraft}, \citenamefont
  {Krin}, \citenamefont {Patterson}, \citenamefont {Doyle},\ and\ \citenamefont
  {Schnell}}]{Shubert2015}%
  \BibitemOpen
  \bibfield  {author} {\bibinfo {author} {\bibfnamefont {V.~A.}\ \bibnamefont
  {Shubert}}, \bibinfo {author} {\bibfnamefont {D.}~\bibnamefont {Schmitz}},
  \bibinfo {author} {\bibfnamefont {C.}~\bibnamefont {Medcraft}}, \bibinfo
  {author} {\bibfnamefont {A.}~\bibnamefont {Krin}}, \bibinfo {author}
  {\bibfnamefont {D.}~\bibnamefont {Patterson}}, \bibinfo {author}
  {\bibfnamefont {J.~M.}\ \bibnamefont {Doyle}}, \ and\ \bibinfo {author}
  {\bibfnamefont {M.}~\bibnamefont {Schnell}},\ }\href {\doibase
  10.1063/1.4921833} {\bibfield  {journal} {\bibinfo  {journal} {J. Chem.
  Phys.}\ }\textbf {\bibinfo {volume} {142}},\ \bibinfo {pages} {214201}
  (\bibinfo {year} {2015})}\BibitemShut {NoStop}%
\bibitem [{\citenamefont {Eibenberger}\ \emph {et~al.}(2017)\citenamefont
  {Eibenberger}, \citenamefont {Doyle},\ and\ \citenamefont
  {Patterson}}]{Eibenberger2017}%
  \BibitemOpen
  \bibfield  {author} {\bibinfo {author} {\bibfnamefont {S.}~\bibnamefont
  {Eibenberger}}, \bibinfo {author} {\bibfnamefont {J.}~\bibnamefont {Doyle}},
  \ and\ \bibinfo {author} {\bibfnamefont {D.}~\bibnamefont {Patterson}},\
  }\href {\doibase 10.1103/PhysRevLett.118.123002} {\bibfield  {journal}
  {\bibinfo  {journal} {Phys. Rev. Lett.}\ }\textbf {\bibinfo {volume} {118}},\
  \bibinfo {pages} {123002} (\bibinfo {year} {2017})}\BibitemShut {NoStop}%
\bibitem [{\citenamefont {P\'{e}rez}\ \emph {et~al.}(2017)\citenamefont
  {P\'{e}rez}, \citenamefont {Steber}, \citenamefont {Domingos}, \citenamefont
  {Krin}, \citenamefont {Schmitz},\ and\ \citenamefont {Schnell}}]{Perez2017}%
  \BibitemOpen
  \bibfield  {author} {\bibinfo {author} {\bibfnamefont {C.}~\bibnamefont
  {P\'{e}rez}}, \bibinfo {author} {\bibfnamefont {A.~L.}\ \bibnamefont
  {Steber}}, \bibinfo {author} {\bibfnamefont {S.~R.}\ \bibnamefont
  {Domingos}}, \bibinfo {author} {\bibfnamefont {A.}~\bibnamefont {Krin}},
  \bibinfo {author} {\bibfnamefont {D.}~\bibnamefont {Schmitz}}, \ and\
  \bibinfo {author} {\bibfnamefont {M.}~\bibnamefont {Schnell}},\ }\href
  {\doibase 10.1002/ange.201704901} {\bibfield  {journal} {\bibinfo  {journal}
  {Angew. Chem.}\ }\textbf {\bibinfo {volume} {129}},\ \bibinfo {pages} {12686}
  (\bibinfo {year} {2017})}\BibitemShut {NoStop}%
\bibitem [{\citenamefont {P\'{e}rez}\ \emph {et~al.}(2018)\citenamefont
  {P\'{e}rez}, \citenamefont {Steber}, \citenamefont {Krin},\ and\
  \citenamefont {Schnell}}]{Perez2018}%
  \BibitemOpen
  \bibfield  {author} {\bibinfo {author} {\bibfnamefont {C.}~\bibnamefont
  {P\'{e}rez}}, \bibinfo {author} {\bibfnamefont {A.~L.}\ \bibnamefont
  {Steber}}, \bibinfo {author} {\bibfnamefont {A.}~\bibnamefont {Krin}}, \ and\
  \bibinfo {author} {\bibfnamefont {M.}~\bibnamefont {Schnell}},\ }\href
  {\doibase 10.1021/acs.jpclett.8b01815} {\bibfield  {journal} {\bibinfo
  {journal} {J. Phys. Chem. Lett.}\ }\textbf {\bibinfo {volume} {9}},\ \bibinfo
  {pages} {4539} (\bibinfo {year} {2018})}\BibitemShut {NoStop}%
\bibitem [{\citenamefont {Ye}\ \emph {et~al.}(2018)\citenamefont {Ye},
  \citenamefont {Zhang},\ and\ \citenamefont {Li}}]{Ye2018}%
  \BibitemOpen
  \bibfield  {author} {\bibinfo {author} {\bibfnamefont {C.}~\bibnamefont
  {Ye}}, \bibinfo {author} {\bibfnamefont {Q.}~\bibnamefont {Zhang}}, \ and\
  \bibinfo {author} {\bibfnamefont {Y.}~\bibnamefont {Li}},\ }\href {\doibase
  10.1103/PhysRevA.98.063401} {\bibfield  {journal} {\bibinfo  {journal} {Phys.
  Rev. A}\ }\textbf {\bibinfo {volume} {98}},\ \bibinfo {pages} {063401}
  (\bibinfo {year} {2018})}\BibitemShut {NoStop}%
\bibitem [{\citenamefont {Vitanov}\ and\ \citenamefont
  {Drewsen}(2019)}]{Vitanov2019}%
  \BibitemOpen
  \bibfield  {author} {\bibinfo {author} {\bibfnamefont {N.~V.}\ \bibnamefont
  {Vitanov}}\ and\ \bibinfo {author} {\bibfnamefont {M.}~\bibnamefont
  {Drewsen}},\ }\href {\doibase 10.1103/PhysRevLett.122.173202} {\bibfield
  {journal} {\bibinfo  {journal} {Phys. Rev. Lett.}\ }\textbf {\bibinfo
  {volume} {122}},\ \bibinfo {pages} {173202} (\bibinfo {year}
  {2019})}\BibitemShut {NoStop}%
\bibitem [{\citenamefont {Wu}\ \emph {et~al.}(2019)\citenamefont {Wu},
  \citenamefont {Wang}, \citenamefont {Song}, \citenamefont {Xia},
  \citenamefont {Su},\ and\ \citenamefont {Jiang}}]{wu2019robust}%
  \BibitemOpen
  \bibfield  {author} {\bibinfo {author} {\bibfnamefont {J.-L.}\ \bibnamefont
  {Wu}}, \bibinfo {author} {\bibfnamefont {Y.}~\bibnamefont {Wang}}, \bibinfo
  {author} {\bibfnamefont {J.}~\bibnamefont {Song}}, \bibinfo {author}
  {\bibfnamefont {Y.}~\bibnamefont {Xia}}, \bibinfo {author} {\bibfnamefont
  {S.-L.}\ \bibnamefont {Su}}, \ and\ \bibinfo {author} {\bibfnamefont {Y.-Y.}\
  \bibnamefont {Jiang}},\ }\href {\doibase 10.1103/PhysRevA.100.043413}
  {\bibfield  {journal} {\bibinfo  {journal} {Phys. Rev. A}\ }\textbf {\bibinfo
  {volume} {100}},\ \bibinfo {pages} {043413} (\bibinfo {year}
  {2019})}\BibitemShut {NoStop}%
\bibitem [{\citenamefont {Ye}\ \emph {et~al.}(2019{\natexlab{b}})\citenamefont
  {Ye}, \citenamefont {Zhang}, \citenamefont {Chen},\ and\ \citenamefont
  {Li}}]{PhysRevA.100.043403}%
  \BibitemOpen
  \bibfield  {author} {\bibinfo {author} {\bibfnamefont {C.}~\bibnamefont
  {Ye}}, \bibinfo {author} {\bibfnamefont {Q.}~\bibnamefont {Zhang}}, \bibinfo
  {author} {\bibfnamefont {Y.-Y.}\ \bibnamefont {Chen}}, \ and\ \bibinfo
  {author} {\bibfnamefont {Y.}~\bibnamefont {Li}},\ }\href {\doibase
  10.1103/PhysRevA.100.043403} {\bibfield  {journal} {\bibinfo  {journal}
  {Phys. Rev. A}\ }\textbf {\bibinfo {volume} {100}},\ \bibinfo {pages}
  {043403} (\bibinfo {year} {2019}{\natexlab{b}})}\BibitemShut {NoStop}%
\bibitem [{\citenamefont {Hirota}(2012)}]{2012PJA8803B-06}%
  \BibitemOpen
  \bibfield  {author} {\bibinfo {author} {\bibfnamefont {E.}~\bibnamefont
  {Hirota}},\ }\href {\doibase 10.2183/pjab.88.120} {\bibfield  {journal}
  {\bibinfo  {journal} {Proc. Jpn. Acad. B}\ }\textbf {\bibinfo {volume}
  {88}},\ \bibinfo {pages} {120} (\bibinfo {year} {2012})}\BibitemShut
  {NoStop}%
\bibitem [{\citenamefont {Glaser}\ \emph {et~al.}(2015)\citenamefont {Glaser},
  \citenamefont {Boscain}, \citenamefont {Calarco}, \citenamefont {Koch},
  \citenamefont {K{\"o}ckenberger}, \citenamefont {Kosloff}, \citenamefont
  {Kuprov}, \citenamefont {Luy}, \citenamefont {Schirmer}, \citenamefont
  {Schulte-Herbr{\"u}ggen}, \citenamefont {Sugny},\ and\ \citenamefont
  {Wilhelm}}]{Glaser2015}%
  \BibitemOpen
  \bibfield  {author} {\bibinfo {author} {\bibfnamefont {S.~J.}\ \bibnamefont
  {Glaser}}, \bibinfo {author} {\bibfnamefont {U.}~\bibnamefont {Boscain}},
  \bibinfo {author} {\bibfnamefont {T.}~\bibnamefont {Calarco}}, \bibinfo
  {author} {\bibfnamefont {C.~P.}\ \bibnamefont {Koch}}, \bibinfo {author}
  {\bibfnamefont {W.}~\bibnamefont {K{\"o}ckenberger}}, \bibinfo {author}
  {\bibfnamefont {R.}~\bibnamefont {Kosloff}}, \bibinfo {author} {\bibfnamefont
  {I.}~\bibnamefont {Kuprov}}, \bibinfo {author} {\bibfnamefont
  {B.}~\bibnamefont {Luy}}, \bibinfo {author} {\bibfnamefont {S.}~\bibnamefont
  {Schirmer}}, \bibinfo {author} {\bibfnamefont {T.}~\bibnamefont
  {Schulte-Herbr{\"u}ggen}}, \bibinfo {author} {\bibfnamefont {D.}~\bibnamefont
  {Sugny}}, \ and\ \bibinfo {author} {\bibfnamefont {F.~K.}\ \bibnamefont
  {Wilhelm}},\ }\href {\doibase 10.1140/epjd/e2015-60464-1} {\bibfield
  {journal} {\bibinfo  {journal} {Eur. Phys. J. D}\ }\textbf {\bibinfo {volume}
  {69}},\ \bibinfo {pages} {279} (\bibinfo {year} {2015})}\BibitemShut
  {NoStop}%
\bibitem [{\citenamefont {Liu}\ \emph {et~al.}(2019)\citenamefont {Liu},
  \citenamefont {Song}, \citenamefont {Xue}, \citenamefont {Wang},\ and\
  \citenamefont {Yung}}]{PhysRevLett.123.100501}%
  \BibitemOpen
  \bibfield  {author} {\bibinfo {author} {\bibfnamefont {B.-J.}\ \bibnamefont
  {Liu}}, \bibinfo {author} {\bibfnamefont {X.-K.}\ \bibnamefont {Song}},
  \bibinfo {author} {\bibfnamefont {Z.-Y.}\ \bibnamefont {Xue}}, \bibinfo
  {author} {\bibfnamefont {X.}~\bibnamefont {Wang}}, \ and\ \bibinfo {author}
  {\bibfnamefont {M.-H.}\ \bibnamefont {Yung}},\ }\href {\doibase
  10.1103/PhysRevLett.123.100501} {\bibfield  {journal} {\bibinfo  {journal}
  {Phys. Rev. Lett.}\ }\textbf {\bibinfo {volume} {123}},\ \bibinfo {pages}
  {100501} (\bibinfo {year} {2019})}\BibitemShut {NoStop}%
\bibitem [{\citenamefont {Wu}\ and\ \citenamefont {Su}(2019)}]{Wu_2019}%
  \BibitemOpen
  \bibfield  {author} {\bibinfo {author} {\bibfnamefont {J.~L.}\ \bibnamefont
  {Wu}}\ and\ \bibinfo {author} {\bibfnamefont {S.~L.}\ \bibnamefont {Su}},\
  }\href {\doibase 10.1088/1751-8121/ab2a92} {\bibfield  {journal} {\bibinfo
  {journal} {J. Phys. A: Math. Theor.}\ }\textbf {\bibinfo {volume} {52}},\
  \bibinfo {pages} {335301} (\bibinfo {year} {2019})}\BibitemShut {NoStop}%
\bibitem [{\citenamefont {Powis}(2008)}]{doi:10.1002/9780470259474.ch5}%
  \BibitemOpen
  \bibfield  {author} {\bibinfo {author} {\bibfnamefont {I.}~\bibnamefont
  {Powis}},\ }\href {\doibase 10.1002/9780470259474.ch5} {\bibfield  {journal}
  {\bibinfo  {journal} {Adv. Chem. Phys.}\ }\textbf {\bibinfo {volume} {138}},\
  \bibinfo {pages} {267} (\bibinfo {year} {2008})}\BibitemShut {NoStop}%
\bibitem [{\citenamefont {Zhang}\ \emph {et~al.}(2013)\citenamefont {Zhang},
  \citenamefont {Shim}, \citenamefont {Niemeyer}, \citenamefont {Taniguchi},
  \citenamefont {Teraji}, \citenamefont {Abe}, \citenamefont {Onoda},
  \citenamefont {Yamamoto}, \citenamefont {Ohshima}, \citenamefont {Isoya},\
  and\ \citenamefont {Suter}}]{PhysRevLett.110.240501}%
  \BibitemOpen
  \bibfield  {author} {\bibinfo {author} {\bibfnamefont {J.}~\bibnamefont
  {Zhang}}, \bibinfo {author} {\bibfnamefont {J.~H.}\ \bibnamefont {Shim}},
  \bibinfo {author} {\bibfnamefont {I.}~\bibnamefont {Niemeyer}}, \bibinfo
  {author} {\bibfnamefont {T.}~\bibnamefont {Taniguchi}}, \bibinfo {author}
  {\bibfnamefont {T.}~\bibnamefont {Teraji}}, \bibinfo {author} {\bibfnamefont
  {H.}~\bibnamefont {Abe}}, \bibinfo {author} {\bibfnamefont {S.}~\bibnamefont
  {Onoda}}, \bibinfo {author} {\bibfnamefont {T.}~\bibnamefont {Yamamoto}},
  \bibinfo {author} {\bibfnamefont {T.}~\bibnamefont {Ohshima}}, \bibinfo
  {author} {\bibfnamefont {J.}~\bibnamefont {Isoya}}, \ and\ \bibinfo {author}
  {\bibfnamefont {D.}~\bibnamefont {Suter}},\ }\href {\doibase
  10.1103/PhysRevLett.110.240501} {\bibfield  {journal} {\bibinfo  {journal}
  {Phys. Rev. Lett.}\ }\textbf {\bibinfo {volume} {110}},\ \bibinfo {pages}
  {240501} (\bibinfo {year} {2013})}\BibitemShut {NoStop}%
\bibitem [{\citenamefont {Zhou}\ \emph {et~al.}(2017)\citenamefont {Zhou},
  \citenamefont {Baksic}, \citenamefont {Ribeiro}, \citenamefont {Yale},
  \citenamefont {Heremans}, \citenamefont {Jerger}, \citenamefont {Auer},
  \citenamefont {Burkard}, \citenamefont {Clerk},\ and\ \citenamefont
  {Awschalom}}]{BBZhou}%
  \BibitemOpen
  \bibfield  {author} {\bibinfo {author} {\bibfnamefont {B.}~\bibnamefont
  {Zhou}}, \bibinfo {author} {\bibfnamefont {A.}~\bibnamefont {Baksic}},
  \bibinfo {author} {\bibfnamefont {H.}~\bibnamefont {Ribeiro}}, \bibinfo
  {author} {\bibfnamefont {C.}~\bibnamefont {Yale}}, \bibinfo {author}
  {\bibfnamefont {F.}~\bibnamefont {Heremans}}, \bibinfo {author}
  {\bibfnamefont {P.}~\bibnamefont {Jerger}}, \bibinfo {author} {\bibfnamefont
  {A.}~\bibnamefont {Auer}}, \bibinfo {author} {\bibfnamefont {G.}~\bibnamefont
  {Burkard}}, \bibinfo {author} {\bibfnamefont {A.}~\bibnamefont {Clerk}}, \
  and\ \bibinfo {author} {\bibfnamefont {D.}~\bibnamefont {Awschalom}},\ }\href
  {\doibase 10.1038/nphys3967} {\bibfield  {journal} {\bibinfo  {journal} {Nat.
  Phys.}\ }\textbf {\bibinfo {volume} {13}},\ \bibinfo {pages} {330} (\bibinfo
  {year} {2017})}\BibitemShut {NoStop}%
\bibitem [{\citenamefont {H\o{}jbjerre}\ \emph {et~al.}(2008)\citenamefont
  {H\o{}jbjerre}, \citenamefont {Offenberg}, \citenamefont {Bisgaard},
  \citenamefont {Stapelfeldt}, \citenamefont {Staanum}, \citenamefont
  {Mortensen},\ and\ \citenamefont {Drewsen}}]{PhysRevA.77.030702}%
  \BibitemOpen
  \bibfield  {author} {\bibinfo {author} {\bibfnamefont {K.}~\bibnamefont
  {H\o{}jbjerre}}, \bibinfo {author} {\bibfnamefont {D.}~\bibnamefont
  {Offenberg}}, \bibinfo {author} {\bibfnamefont {C.~Z.}\ \bibnamefont
  {Bisgaard}}, \bibinfo {author} {\bibfnamefont {H.}~\bibnamefont
  {Stapelfeldt}}, \bibinfo {author} {\bibfnamefont {P.~F.}\ \bibnamefont
  {Staanum}}, \bibinfo {author} {\bibfnamefont {A.}~\bibnamefont {Mortensen}},
  \ and\ \bibinfo {author} {\bibfnamefont {M.}~\bibnamefont {Drewsen}},\ }\href
  {\doibase 10.1103/PhysRevA.77.030702} {\bibfield  {journal} {\bibinfo
  {journal} {Phys. Rev. A}\ }\textbf {\bibinfo {volume} {77}},\ \bibinfo
  {pages} {030702(R)} (\bibinfo {year} {2008})}\BibitemShut {NoStop}%
\bibitem [{\citenamefont {Balle}\ and\ \citenamefont
  {Flygare}(1981)}]{doi:10.1063/1.1136443}%
  \BibitemOpen
  \bibfield  {author} {\bibinfo {author} {\bibfnamefont {T.~J.}\ \bibnamefont
  {Balle}}\ and\ \bibinfo {author} {\bibfnamefont {W.~H.}\ \bibnamefont
  {Flygare}},\ }\href {\doibase 10.1063/1.1136443} {\bibfield  {journal}
  {\bibinfo  {journal} {Rev. Sci. Instrum.}\ }\textbf {\bibinfo {volume}
  {52}},\ \bibinfo {pages} {33} (\bibinfo {year} {1981})}\BibitemShut {NoStop}%
\bibitem [{\citenamefont {Grabow}(2013)}]{doi:10.1002/anie.201307159}%
  \BibitemOpen
  \bibfield  {author} {\bibinfo {author} {\bibfnamefont {J.-U.}\ \bibnamefont
  {Grabow}},\ }\href {\doibase 10.1002/anie.201307159} {\bibfield  {journal}
  {\bibinfo  {journal} {Angew. Chem. Int. Ed.}\ }\textbf {\bibinfo {volume}
  {52}},\ \bibinfo {pages} {11698} (\bibinfo {year} {2013})}\BibitemShut
  {NoStop}%
\end{thebibliography}%
\end{document}